# Revisiting Citizen Science Through the Lens of Hybrid Intelligence

Janet Rafner[a], Miroslav Gajdacz[a], Gitte Kragh[a], Arthur Hjorth[a], Anna Gander[b], Blanka Palfi[a], Aleks Berditchevskaia[c], François Grey[d], Kobi Gal[e], Avi Segal[e], Mike Walmsley[f], Josh Aaron Miller[g], Dominik Dellerman[h], Muki Haklay[i], Pietro Michelucci[j], Jacob Sherson[a]

[a] *Center for Hybrid Intelligence, Denmark*
[b] *University of Gothenburg, Sweden*
[c] *Centre for Collective Intelligence Design, United Kingdom*
[d] *Citizen Cyberlab*
[e] *Ben Gurion University, Israel*
[f] *University of Oxford, England*
[g] *Northwestern University, USA*
[h] *Vencortex*
[i] *University College London, England*
[j] *Human Computation Institute, USA*

## Abstract

Artificial Intelligence (AI) can augment and sometimes even replace human cognition. Inspired by efforts to value human agency alongside productivity, we discuss the benefits of solving Citizen Science (CS) tasks with Hybrid Intelligence (HI), a synergetic mixture of human and artificial intelligence. Currently there is no clear framework or methodology on how to create such an effective mixture. Due to the unique participant-centered set of values and the abundance of tasks drawing upon both human common sense and complex 21st century skills, we believe that the field of CS offers an invaluable testbed for the development of HI and human-centered AI of the 21st century, while benefiting CS as well. In order to investigate this potential, we first relate CS to adjacent computational disciplines. Then, we demonstrate that CS projects can be grouped according to their potential for HI-enhancement by examining two key dimensions: the level of digitization and the amount of knowledge or experience required for participation. Finally, we propose a framework for types of human-AI interaction in CS based on established criteria of HI. This "HI lens" provides the CS community with an overview of several ways to utilize the combination of AI and human intelligence in their projects. It also allows the AI community to gain ideas on how developing AI in CS projects can further their own field.

# 1. Introduction

Citizen Science (CS) attracts hundreds of thousands of participants to research projects that utilize human problem-solving for tasks of varying complexity (Heck et al. 2018; Huang et al. 2018). This collective intelligence of participants has contributed to many scientific discoveries, for example in protein folding (Cooper et al. 2010, Koepnick et al. 2019), galaxy morphology identification (Lintott et al. 2011, Masters and Galaxy Zoo Team 2019), mapping of neurons in the brain (Kim et al. 2014), heuristics for solving quantum physics challenges (Jensen et al., 2021), tracking changes in biodiversity and ecology (Lim et al. 2019, Wang et al. 2018), observing and monitoring air pollution (Snik et al. 2014), and transcribing manuscript collections (Causer and Terras 2014). Some of these achievements have been made in part by participants going beyond the narrow task given to them, digging further into the data to understand and interpret what they are seeing (Lintott 2019).

Scientific research often benefits from CS in ways that go beyond "free" labor. For example, researchers use CS when they are unable to collect the necessary data by themselves (Wyler et al. 2016), need specific expertise from the general public to help solve a problem (Danielsen et al. 2018), datasets are too large or complicated for the researchers to process with their given technology and resources (Das et al. 2019, Fortson et al. 2011, Nugent, 2019), or the degrees of freedom of a system results in nearly infinite possible candidate solutions to be explored (Jensen et al. 2021, Koepnick et al. 2019). Simply put, CS projects enhance scientific research by tapping into the collective cognitive and labor resources of the general public.

In a similar vein, Artificial Intelligence (AI) provides value in many scientific disciplines, making it possible for researchers to work with larger amounts of data, or detect patterns that would be hidden to the human eye or more reductionist statistical methods. Broadly, AI is being used to solve a wide variety of problems ranging from game challenges (Silver et al. 2016) to corporate applications (Eager et al. 2020, Li et al. 2017). Due to the quantitative character of natural sciences they are particularly well suited for Machine Learning applications (ML; a sub-field of AI) across disciplines like physics (Bohrdt et al 2019, Dalgaard et al. 2020) astronomy (Godines et al. 2019, Ormiston et al. 2020), biology (Senior et al. 2020), ecology (Leoni et al. 2020), and geoscience (Wang et al. 2019) to name a few. Cognitive science (Lake et al. 2015) and social sciences (Chen et al. 2018, Hindman 2015) also employ ML methods with increasing frequency. Furthering our understanding of AI will in turn enhance future scientific progress.

Despite the success of AI, a growing part of the AI community has realized that many tasks can still only achieve the required quality and reliability with some form of human-in-the-loop (Benedikt et al. 2020, Zanzotto 2019) input to the particular task. This could be either live or prerecorded human input as we show in this paper. Purely deep learning-based approaches often identify particular patterns in the training data that are not robust enough to solve many real-world problems in noisy, unpredictable and varying environments (Heaven 2019, Marcus, 2018). Also, despite the spectacular successes of AlphaZero to "teach itself" to play Chess and Go (Silver et al. 2018), the DeepMind team had to employ extensive learning from human gameplay in order for it to succeed in the complex and dynamic multi-actor environment of Starcraft (Vinyals et al 2019). The limitations of current "self-learning" AI is highlighted by the, perhaps surprising, fact that AlphaZero has so far hardly found application beyond the realm of games (Dalgaard et al. 2020, Tomašev et al. 2020). One goal of AI research is therefore to better understand what makes tasks unsolvable by current algorithms, and find alternative or complementary methods.

One emerging response to the failure of achieving autonomous operation is to develop hybrid solutions bringing the human more intimately into the loop, optimally combining the information processing capabilities of both humans and machines (Michelucci and Dickinson 2016, Dellermann et al. 2019, Christiano et al. 2017). Many such advanced approaches focus on capturing failures of the stand-alone AI system by querying humans for feedback about a certain selection of the AI predictions (Kamar and Manikonda 2017, Nushi et al. 2018). This becomes crucial in high stake applications, such as medical diagnostics (Holzinger 2016, Wilder et al. 2020). These approaches have also started finding their way into research contexts such as the optimization of complex dynamical systems (Baltz et al. 2017). In the latter study, subjective human expert opinion was used to choose one of the viable actions proposed by the AI, which led to superior results in fusion research compared to the case of pure machine control.

The field of CS is ideal for developing hybrid intelligence interactions for two distinct reasons. First, although all fields of science revolve around human problem-solving, the human computation being performed is often defined implicitly through the tacit domain-specific experience of the involved experts. In contrast, the field of CS specializes in explicitly transforming conventional research challenges into tasks tapping into the problem-solving abilities and collective intelligence of the general public. Second, the full long-term value of hybrid interactions may not always be immediately apparent because developing interfaces to optimally support human creativity is very challenging. Therefore, commercial applications may tend to focus more narrowly on short-term efficiency maximization using shallower, but predictable human involvement. In contrast, the field of CS is fueled both by a desire to solve concrete tasks but also to generate intrinsic value for the participants through as deep and meaningful involvement in the projects as possible and as participants wish. Additionally, although AI methodologies are starting to be applied in CS projects, little attention has been given to bi-directional human-computer interactions. We argue that the combination of these practical and value-based considerations make CS particularly well-suited to develop approaches combining human and artificial intelligence, i.e. hybrid intelligence, into concrete projects that will benefit the field of AI, the CS projects and participants, and science and society at large.

The concept of Hybrid Intelligence (HI) referred to above has been rather loosely defined in many variations (Akata et al. 2020, Lasecki 2019, Prakash and Mathewson 2020). Here we adhere to the operational HI definition in terms of three criteria put forward by Dellermann et al. 2019:
- *Collectiveness:* the human and AI are solving the task collectively towards a system-level goal. Sub-goals of individual agents might be different from the system-level goal

- **Solution superiority:** the sociotechnical system achieves results superior to that one of the individual agents (human or AI)
- **Mutual learning:** the system improves over time, both as a whole and also each single component (human and AI)

As we show in this paper, these three criteria of HI can serve as a lens for classifying the different interaction schemes between participants and AI in CS projects.

The aim of this paper is to identify which processes and outcomes within CS could benefit from combining AI and human intelligence, though many of the observations can be generalized to use-cases outside of CS. For the CS community, this paper provides a visualisation of how AI can support their projects. For AI researchers, this work highlights the opportunity CS presents to engage with real-world data sets and explore new AI methods and applications. In particular, CS projects are a fertile ground for quantitatively and qualitatively studying human 21st century skills such as creativity, hierarchical thinking and common sense (Jensen et al. 2020) which is a current roadblock in developing more robust AI (Marcus 2018). For both, there are opportunities for interdisciplinary and transdisciplinary collaborations.

In order to investigate which types of AI can and should be integrated into CS projects, we first relate CS to adjacent computational disciplines. A conceptual mapping of terms is necessary to effectively exploit the extensive insights from each research tradition in the interdisciplinary endeavor towards optimal human-machine problem-solving. Then, we examine several projects and their potential for AI-enhancement through two key dimensions: the degree of digitization and the accessibility to make a scientific contribution. We do this to concretely illustrate which types of CS tasks are ideally suited for which types of machine support. Finally, we present a framework for types of human-AI interaction in CS based on established criteria of HI. This framework identifies and categorises *the ways* AI can augment the process of solving CS tasks.

## 2. Types of Computation, Emergent Intelligence and CS

In order to find optimal ways for AI to enhance CS we start by exploring the axes of *types of computation* (biological, human, and machine) and *number of agents*. Each axis has an *emergent intelligence* (Bonabeau et al. 1999) associated with it: *hybrid intelligence* and *collective intelligence* respectively. We provide an overview table on the terms used (Table 1), followed by a diagram illustrating their relationships (Fig 1).

| Computation<br>General information processing | Artificial Intelligence<br>Form of machine computation applied to tasks that have traditionally required human intelligence (Bellman, 1978) |
|---|---|
| **Biological computation** (Mitchell, 2011)<br>● A form of computation<br>● Performed by living organisms<br>● parallel, stochastic, with no clean notion of a mapping between "inputs" and "outputs"<br>**Human computation** (Michelucci, 2013)<br>● A form of biological computation<br>● Performed by an individual human, groups of humans, and can be mediated by machines<br>● Human intelligence is always used as humans are assumed to always act with 'intelligence'<br>**Machine computation** (Copeland, 1997)<br>● Information processing through clearly defined rules (algorithms), done by machines/computers which may not always be 'intelligent' but also includes AI | **Machine Learning (ML) overview**<br>● Suite of methods used in AI<br>● Based on algorithms whose function is not fully determined by a program but rather acquired via a learning session called *training*<br>● Example input-output data pairs presented to the algorithm while parameters (collectively called '*the model*') of the algorithm are empirically adjusted so that the input produces the desired output<br>● Once the model is trained, parameters are fixed and the algorithm can be *executed* on novel inputs to provide predictions or decisions<br>**ML type 1: Supervised learning**<br>● *Objective*: predicting a specific class label classification) or a parameter value (regression) for any given input<br>● *Training* contains ground truth labels assigned to a set of individual input examples which is usually obtained by human labelling or resulting from experimentation or simulation |
| **Emergent Intelligence**<br>Emerges when multiple agents (human or machine) collaborate on a common goal | **ML type 2: Unsupervised learning**<br>● *Objective*: classify input samples without knowledge of class labels using a measure of similarity, typically calculated as distances in many dimensional space<br>● Includes dimensionality reduction where prominent features or representations are extracted based on their ability to to distinguish the input samples |
| **Collective Intelligence** (Lévy & Bononno, 1997; Russell, 1995)<br>● Associated with collaboration or competition of **many** individual agents (human **AND/OR** machine)<br>● Emerges with 3 or more agents (Woolly et al., 2010)<br>**Hybrid Intelligence** (Dellermann et al., 2019)<br>● Arises in systems combining human agents **AND** AI<br>● Superior results are achieved than either the human or machine could have accomplished separately<br>● Both the human(s) and AI(s) continuously improve by learning from each other | **ML type 3: Reinforcement learning**<br>● *Objective*: optimize behavior of an AI agent which reacts often repeatedly, to the state of the environment<br>● Agent received feedback on its action from a reward function<br>● The environment can be real (as in robotics), a closed digital model like a game world, or a simulation of some process |

**Table 1:** Overview of types of computation, emergent intelligence, and artificial intelligence that are referred to throughout the paper.

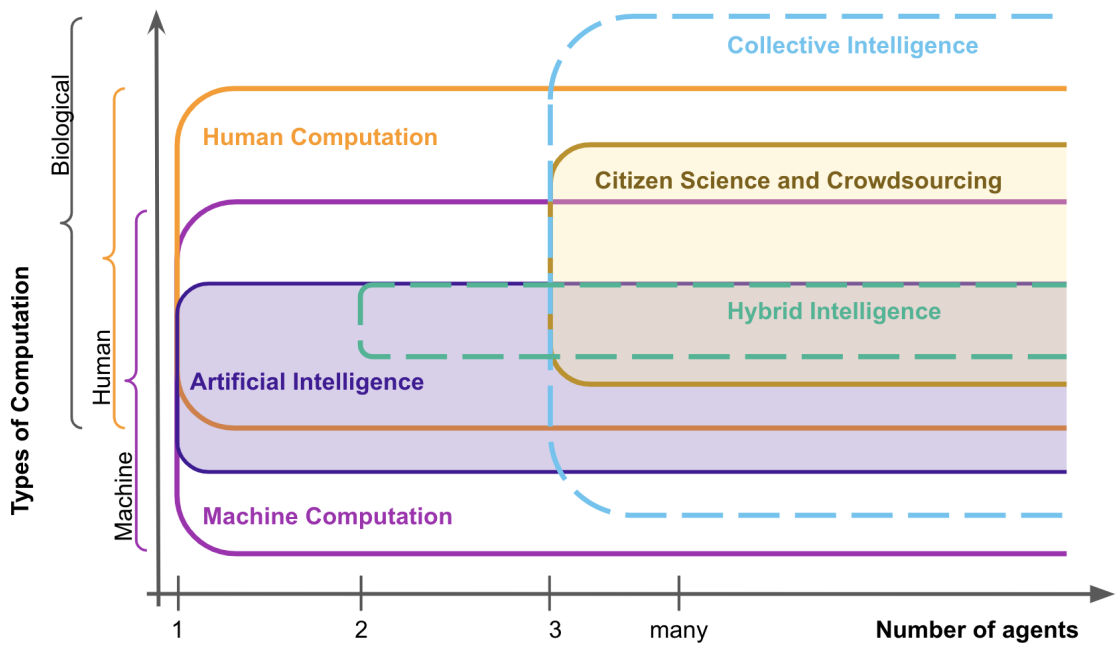

**Figure 1:** The diagram illustrates the relationship of Hybrid Intelligence and Citizen Science in the reference frame of mixed-agent computation (y-axis), moving from machine to human and finally to general biological computation, and agent (biological individuals or machines) count (x-axis) moving from one agent to many.

HI is a subset of the overlap between Human Computation and AI. As HI can be achieved with only two agents, it lies partially outside of Collective Intelligence which requires at least three agents (Woolley et al. 2010). Apart from collectiveness and solution superiority, HI poses a rather strict requirement of mutual learning, which explains the substantial overlap between CS and AI as well as an overlap of AI and Human Computation beyond the area of HI. Since few projects today achieve all three HI requirements, the size of the HI field on fig 1 is overrepresented. However, HI's importance will likely accelerate as algorithmic development increasingly focuses on human-centered AI (Auernhammer 2020). For the rest of this paper, we focus on the overlap between CS (fig. 1, yellow box) and Al (fig. 1, purple box), and discuss the characteristics which make projects lie inside or outside the realm of HI.

## 3. Task Characteristics in Citizen Science

We now discuss the ***degree of digitization*** and the ***accessibility to scientific contributions*** of CS projects to determine which tasks are ideally suited for which types of machine support. To date, well-known CS typologies have focused on participant contributions to different parts of the research process such as collecting data (Bonney et al. 2009b, 2016, Paul et al. 2014), forming hypotheses (e.g. Bonney et al. 2009a, Haklay 2013), or generating knowledge in the project (Schäfer and Kieslinger 2016). Wiggins & Crowston (2011) were the first to take into account what they termed 'virtuality', i.e. digitization, in which projects are characterized as either online or offline. As nearly all CS projects trend towards including some digital components, in the following we show how adding granularity to the ***degree of digitization*** lends valuable insights into the potential forms of human-machine interactions in CS tasks. The degree of digitization is closely related to the choice of AI to optimally support CS participants in solving a task. To further understand the type of support needed and thereby the choice of AI, we also explore a second CS task characteristic: ***accessibility to contribution***, which we elaborate on below. Note, even though a task may routinely be solved by participants, it should not be taken as a sign of computational simplicity, since tasks easily completed by humans (e.g., pattern recognition) can be quite challenging for AI. This CS task mapping may lead to increased appreciation of the multitude of human-processing going on in CS projects that are still far from being automatable in any foreseeable future.

**Digitization degree of the CS task**

We propose a granular description and classification of projects based on three different categories of task digitization: optimization tasks, annotation tasks, and physical tasks.

***Physical Tasks*** require participants to perform non-digital actions, to acquire data such as birdwatching. In these tasks, the participant needs to continually (audio-)visually survey the environment and/or consider the suitability of deploying a sensor for recording data (Camprodon et al. 2019, Van Horn et al. 2018, D'Hondt, Stevens and Jacobs 2013, Cochran et al. 2009). The machine analogy of the data collection task would be robotics and smart sensors. In smart sensors, the raw measurement data is processed locally in the hardware before being passed to a central data storage for further processing (Posey, B. n.d.). We are currently unaware of any CS projects that employ smart sensors.

***Annotation tasks*** are solved via a digital platform, but require subject-specific or disciplinary knowledge, even if at a layperson level. Thus one cannot score the participants' input objectively. Instead, the annotation is consensus-based (absence of a ground truth). The elements to be annotated are often images or audio recordings or transcriptions (Lintott et al. 2008, Nugent 2019, Tinati et al. 2017, Causer and Terras 2014). The annotated data can be used to train ML classification models which fall into the paradigm of supervised learning (SL).

***Optimization tasks*** are completely digital and are related to systems which can be described with a self-contained mathematical model (Curtis 2015, Lee et al. 2014, Jensen 2021, Wootton 2017). By self-contained model, we mean a task that can be unambiguously and automatically evaluated (scored) in terms of how well a candidate solution solves the problem without any further human input. These are problems that can often - in theory - be solved purely by machine computation, but in practice may become intractable due to high complexity of the solution space. Naturally, these tasks lend themselves well to ML methods related to optimization in complex spaces, such as reinforcement learning (RL).

The ***degree of digitization*** allows for a rough categorization of the possible AI-contribution as robotics/smart sensors, SL, or RL respectively. One might naively expect that the pinnacle of human contribution is to solve complex mathematical problems (high degree of digitization tasks). However, in many ways the robotics/ smart sensors capable of assisting or replacing human volunteers in the real world for physical tasks is a much more difficult problem, considering current robotics are comparably less advanced than the state-of-the-art RL technologies (Vinyals et al. 2019, Dalgaard et al. 2020). This clearly demonstrates that the degree of digitization should not be mistaken for an axis of increasing cognitive complexity. We therefore posit that a systematic comparison with modern computational capabilities will lead to increased understanding and appreciation of the multitudes of tasks that human volunteers perform, and how that labor works alongside technology and ML.

The x-axis of Figure 2 plots projects according to the degree of digitization. As illustrated, there may be projects exhibiting a mixture of features from two categories. Another boundary case is CS remote optimization of concrete experiments (Heck et al. 2018), which would clearly be amenable to RL treatment but requires execution of a real-world experiment in order to evaluate the quality of any given user-specified candidate solution and therefore does not have a self-contained model.

Finally, we note that the ML technique of Unsupervised Learning is absent here because it is a data analysis technique that can in principle be applied to any data set across the categories. In the following we illustrate the relevance of these task characteristics through several specific CS projects (see appendix A for all considered projects).

**Accessibility to make a scientific contribution**

Whereas the degree of digitization allows for a rough categorization in terms of the potentially applicable forms of AI, it does not explain why some tasks within each category are easy for most participants and some can only be solved by a minority. To address this question, we propose the term, ***accessibility to contribution***, which we define as: "the likelihood that an average layperson (assuming there are no impediments to participation, e.g., physical, socio-cultural, financial, or technological) would make a scientific contribution to the particular project." To elaborate on this term, we present 9 examples, a subset of CS projects reviewed, and discuss what it takes for a participant to make a scientific contribution. We propose a spectrum along which projects can be ordered to signify broader or more limited accessibility.

*Optimization tasks*

> **Quantum Moves 2** is a real-time dynamics control game designed to tap into a player's *intuition* of water sloshing in a glass as they move an atom through a 2-dimensional space over the span of a few seconds. Apart from investigating the value of each individual human input, there is an emphasis on understanding the aggregated collective input to gain understanding of the generic intuition-driven strategies (Jensen et al. 2021). The data analysis in Quantum Moves 2 builds on a bulk analysis of all player data and thus these heuristics are gleaned from all player data. *broad accessibility*

> **Foldit** is a puzzle-type game designed to visualize proteins in three dimensions, and lets participants spend as long as they need to slowly and (semi-)systematically search through a complex parameter landscape as they attempt to find the best folding pattern for a specific protein (Cooper et al. 2010). Reported results focus on the small subset of participants that arrive at uniquely useful solutions (Eiben et al. 2012, Khatib et al. 2011). *medium-limited accessibility*

> **Decodoku** is designed for participants to solve sudoku-like quantum computing challenges without a time limit (Wootton 2017). Data are only collected in the form of written reports emailed to the scientists where participants not only have to come up with useful strategies but also be able reflect on and verbalize their strategies. *limited accessibility*

*Annotation tasks*

> **Stall Catchers** is designed to facilitate analysis of data related to Alzheimer's research. Participants are presented with few-second video clips of blood vessels from the brain of mice affected with Alzheimer's. Through analyzing the movement of blood cells in a target area determined by the game, they classify images as either flowing or stalled, and mark the precise location of stalls on the images. The puzzles require non-domain specific skills and can be solved with minimal domain knowledge (Nugent 2019). *broad accessibility*

In **Galaxy Zoo** participants classify images of galaxies according to a series of questions (Lintott et al. 2008). Some questions are approachable with minimal domain knowledge (e.g., "Does this galaxy have spiral arms?") while others benefit from experience (e.g., "Is there anything odd?"). Examples and illustrative icons help teach new participants how to participate *medium-broad accessibility*

**Scribes of Cairo Geniza** is a transcription project, where participants are presented with images of historic text fragments in Hebrew and Arabic and transcribe it one line at a time using an online program. Participation requires specialized training and/or prior knowledge when dealing with specialized objects due to language requirements (Scribes of the Cairo Geniza n.d.). *limited accessibility*

*Physical Tasks*

**Quake-Catcher Network** is a real-time motion sensing network of computers for earthquake monitoring. Participants download the software and purchase a USB sensor device, which records seismological waves while the software algorithmically determines waves outside the normal range, and sends them back to the project server. Participation, apart from the initial setup, does not require active action or skills of the participant (Cochran et al. 2009). *broad accessibility*

**iNaturalist** is an online social network, where participants can share biodiversity information by recording observations of organisms or their traces (nests, tracks etc.). Users can add identifications to these observations and an automated species identification algorithm is also used on the platform. Participation requires none to extensive domain-specific skills, depending on whether the user wants to also perform identification tasks. Observations can be used to monitor organisms at various locations. (iNaturalist 2021). *broad - medium accessibility*

**UK Butterfly Monitoring Scheme (UKBMS)** is a recording protocol used to record data on the butterfly population. Participants walk 1-2 km routes weekly at specific times of the day, in specific weather conditions from spring to fall multiple years in a row. The task is to record measurements on e.g., weather, habitat, and the number of different butterfly species on recording forms which are submitted weekly on the project website. Significant time investment, prior domain knowledge, and detailed environmental surveying skills are required; participation is limited to the United Kingdom (Dennis et al. 2017). *limited accessibility*

As we see, accessibility to contribution is determined by requirements such as: expertise through training (e.g., animal identification skills), experience (becoming familiar with the task environment and interface, e.g., Foldit), certain cognitive skills (currently understudied in CS). If these factors are properly understood, AI can be used to broaden accessibility by automatically adapting to the diverse needs and skills of participants, facilitating quality of contributions (Anderson-Lee et al. 2016, Walmsley et al., 2020.) and making the task simpler and more enjoyable for participants (). Attempts have been made to optimize the interactions between the volunteers and the scientific tasks of the CS project increasing engagement and optimizing quality of contributions (Sterling 2013). However, without an appropriate underlying framework these specific examples are difficult to generalize, as both CS tasks and the needs of volunteers are diverse.

The accessibility to make a contribution axis highlights a gap in research studying the diverse cognitive skills of participants with respect to the requirements of the scientific task in CS projects. We argue that our categorization allows for joint considerations about cognitive and learning processes of participants as well as possible computational models of AI applicable across a wide range of CS projects. In particular, we demonstrate that across CS projects there exist a class of problems that nearly all participants can contribute to using general human cognitive and motoric abilities (see Fig 1, y axis). The apparent simplicity of these tasks from the human perspective stands in dramatic contrast to the challenge of replicating them with AI technologies, which is one of the grand challenges of the AI field (Marcus 2018, 2020). At the other end of this spectrum lie projects where only a small fraction of participants are able to contribute. Here, understanding of how task learning can be combined with systematic exploration and intuitive leaps remains another grand challenge of AI. Interestingly, most limited accessibility tasks also draw heavily on many of the 21st century skills such as creativity and complex problem solving, which still elude a firm theoretical understanding in the field of psychology and education. Nevertheless, a further analysis of the particular cognitive processes going on in CS projects will be crucial for designing future automated support systems to enhance the contribution of the participants.

In general, the fields of AI and CS would benefit greatly from research unpacking the link between the accessibility of concrete CS tasks and the particular cognitive processes required to complete them. Such an analysis is well beyond the scope of this work, however, we do note that many high accessibility tasks are characterized by intuitive processing and the application of common sense (information processing or physical actions that most participants can do instinctively). Finally, the meta-cognitive aspects in the Decodoku example illustrates that it would be interesting to relate the accessibility level to the emerging concept of co-created CS (Bonney et al. 2009b), in which participants are involved not just in the data gathering phase of the scientific process but also e.g. hypothesis generation, design, and analysis. The algorithmic support of the scientific processes beyond data acquisition could tap into cutting edge AI trends such as unsupervised ML, hierarchical modelling (Menon et al 2017) and generative design (McKnight 2017, Oh et al. 2019).

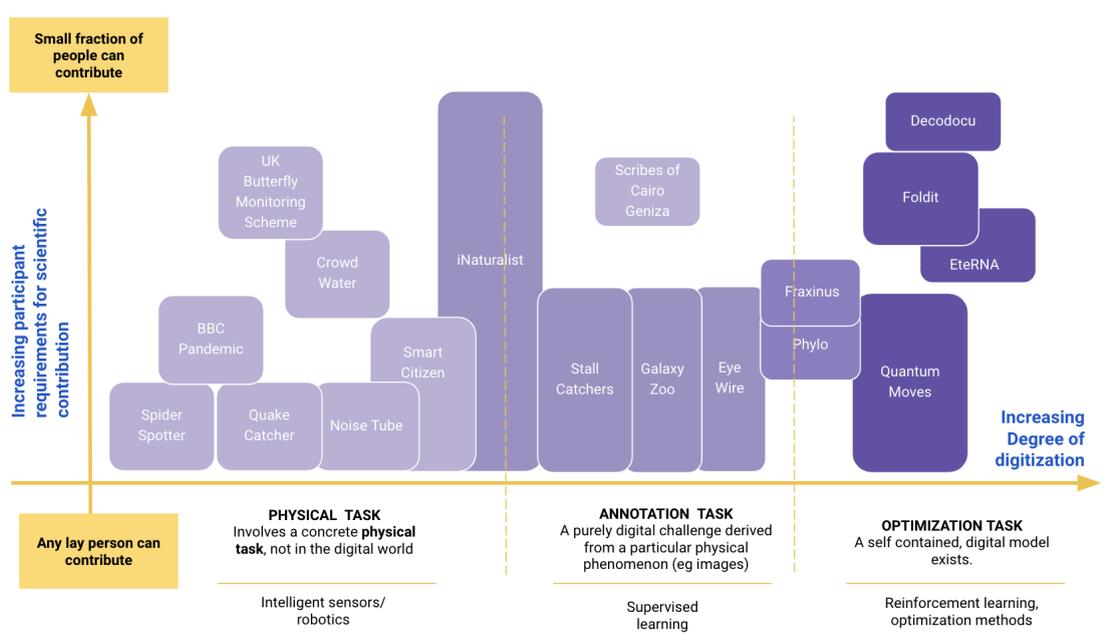

**Figure 2:** Mapping of Citizen Science projects. The x-axis shows an increasing degree of digitization moving from physical tasks (potentially supported by robotics and smart sensors), through annotation tasks (potentially supported by supervised learning methods) to purely mathematical, optimization tasks (potentially supported by reinforcement learning methods). The y-axis represents accessibility to scientific contribution with highly accessible projects at the bottom and projects with extensive requirements on special cognitive traits, expert knowledge, or training at the top.

To summarize, for the CS community, this categorization provides an overview of how different CS projects do or could deploy AI technology as well as an opportunity to reflect on the tasks performed by their participants. For AI researchers, it may allow them to identify methodically suitable collaborations with CS.

## 4. Towards a Framework for Classifying AI Applications in CS Projects From the HI Perspective

Now that we have characterized the types of CS tasks, we turn to identifying and categorizing the ways AI can augment the process of solving these CS tasks. (Ceccaroni et al. 2019) identified three "broad and overlapping" categories for the use of AI in CS. These are: "...[1] *assisting or replacing humans in completing tasks*... enabling tasks traditionally done by people to be partly or completely automated. ...[2] *influencing human behaviour*... e.g., through personalisation and behavioural segmentation, or providing people a means to be comfortable with citizen science and get involved. ... [and 3] having *improved insights* as a result of using AI to enhance data analysis." This categorisation can be useful in distinguishing between the involved groups of humans: the CS participants and the scientists, referring to the first and third category respectively. If, however, we consider "generating improved insights" as part of the CS task (e.g., users perform their part of the task and their input is later aggregated by the AI to yield the desired solution (insight)), we can join the first and third category into one, simply called: ***Assistance in solving the CS task.*** Our criteria for this category is that the AI has to be *directly involved in the solution of the CS task*. Doing so, the AI may either provide a problem-related input to the CS participant during the task, or process the participant's input to achieve or improve the solution.

In other words, one could say that *assistance in solving the CS task* occurs when the *AI is applied inside of the CS task*. Naturally, one would ask what is the contrasting case of *AI being applied outside of the CS task*? Can it be assimilated with the "*influencing human behavior*" AI application category mentioned above? We believe the answer is yes, but in order to make the category more inclusive and to take the CS task as a reference point, we propose to rename it to: **CS content or task selection**. This is to be understood in a broader sense as "intelligent content selection" for the CS participants with the purpose to either incentify a desired behavior (e.g., increase engagement and retention through motivational messages) or to harness the human expertise more efficiently (e.g., selecting tasks where the human input is most valued). In the discussion below, we take the participant-centered view, defining the task as a single instance of a problem.

Given the above definitions, we can already examine the relation of these categories to HI. Revisiting the HI criteria (collectiveness, solution superiority and mutual learning), we see that the category *CS content or task selection* does not meet the criteria of *Collectiveness*, since the AI is *not working on the same task* as the participants. In contrast, the *Assistance in solving the CS task* satisfies the HI Collectiveness criteria per definition. Reversing the argument, the HI criteria of Collectiveness can be used to divide the AI applications in CS into the two above-stated categories. Our HI framework is in no way intended to judge the value of AI applications or diminish the value of intelligent tasks or content selection in CS. Rather, it is intended as a design guide and a mental model in scenarios where the HI scheme can provide solution efficiency and boost participant satisfaction. Now we proceed to the explicit examples from the two AI application categories.

### *Task assistance examples*

In the CS task assistance category, the use of Supervised ML with Deep Neural Networks (DNNs) for image classification is prominent. Here participants are tasked with labeling observations which can then be used for training better AI models, automating the task or parts of it. For example, Galaxy Zoo recently added a Bayesian DNN able to learn from volunteers to classify images of galaxies (Walmsley et. al 2020). Similarly, in (iNaturalist 2021) a DNN model trained on scientific grade data can provide good suggestions of species names or broader taxons such as genera or families for pictures submitted by participants. The participants are also tasked with classification of the images, therefore the human and AI agents are solving the CS task collectively.

To our knowledge RL has not been applied to CS task assistance. However, it holds great potential and initial steps have been taken by integrating simpler, non-learning algorithms directly into a number of CS optimization tasks (Koepnick et al., 2019; Jensen et al., 2020). Even though these algorithms do not learn, they can greatly enhance performance of the human-machine system, achieving solutions superior to that of each agent alone. For example in Foldit, machine agents can assist the participant in the solution of the task in two different ways: first, systematically optimizing a participant's solution with "small tweaks", using an optimization algorithm initiated by the participant (Cooper et al. 2010, pp. 756). Secondly, the Foldit software assists the participant by continuously calculating a variety of assessment criteria for the protein that the participant is currently folding, and thus through continuous feedback providing the participant with additional information (Kleffner et al. 2017, pp. 2765–2766). Similarly, in Quantum Moves 2, players can engage a local optimizer starting from their "hand drawn" solution. It has been demonstrated that in some complex problems, the combination of such human seeding with machine optimisation can provide superior results to that of the individual machine agents (Jensen et al. 2021).

**Content and task selection examples**

AI applications in the CS content and task selection category do not engage in solving the CS task directly, instead they aim at increasing the participants' productivity and/or motivation, which could also be called participant management. This can be done with or without personalizing the selection for individual participants. The Galaxy Zoo *Enhanced mode* (Walmsley et al. 2020) (discussed in the following section) provides players only with tasks which are hard for the AI to solve alone. (See below for a precise definition of the CS task.) In contrast, in personalized task selection, an AI agent takes into account the behavior of the individual participants and directs them toward tasks better suited for their interests/abilities. CS task selection can also include displaying motivational messages, as these may indirectly change the perception of the current task or channel the users to an alternative task or content. For example, (Segal et al. 2018) present their approach, Trajectory Corrected Intervention, in which they used ML to select custom motivational messages for individual users. By doing so, they improve the retention of participants in Galaxy Zoo (Segal et al., in press). Spatharioti et al. (2019) also target participant engagement in the Cartoscope project by making the task and interface more interesting to the participant using an algorithm that adapts task difficulty to the player behavior. Crowston et al. (2020) implemented a Bayesian volunteer learning model to optimize training of newcomers by directing the flow of image classification tasks so that they learn more quickly while meaningfully contributing to the project at the same time. As a last example, Xue et al. (2016) applied an agent behavior model to reduce bias by optimizing incentives to motivate participants to collect data from understudied locations.

With these ever-improving AI technologies comes an abundance of opportunities for CS projects to distribute tasks more efficiently among humans and machines, to support the participants in solving the tasks, and to enhance the humans' experience and consequently enhance their contribution. To optimally take advantage of these opportunities, we need to understand how information is processed differently in relation to solving CS tasks. The next section describes a way to visually represent this information processing in order to underpin the understanding needed.

## Generic Information Flow Diagrams (IFD)

The two previously described types of AI applications in CS, task assistance or selection, differ in the way information is processed. This can be graphically distinguished using *information flow diagrams* (Wintraecken 2012) (fig. 3).

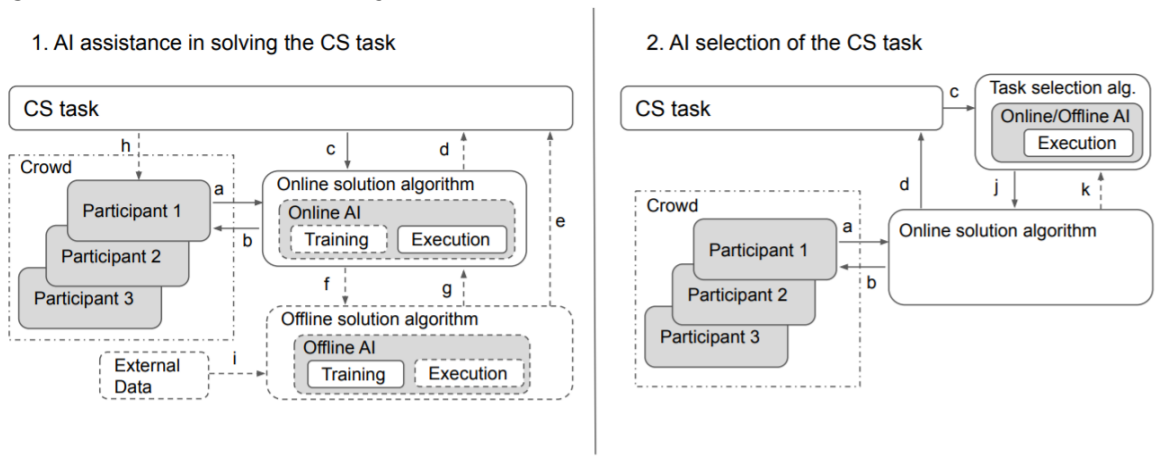

**Figure 3.** Information flow diagrams for the two types of AI applications in Citizen Science (CS) projects: 1. Assistance in solving the CS task; 2. Content or task selection. Arrows denote flow of information (labeled with lowercase letters; referred to in the text); containers denote information sources/receivers, which can also be referred to as agents (machine or human). Agents containing natural or artificial intelligence are marked by grey color fill. Optional components are denoted by a dashed outline. See the following table for explanation of the terms and significance of letters.

| Arrow | Required | Sender | Receiver | Information content |
|---|---|---|---|---|
| a | yes | Participant | OnSA | Participant's input/contribution to CS |
| b | yes | OnSA | Participant | Representation of the task and feedback on participant's actions |
| c | yes | CS task | OnSA | Specification of the task |
| d | no | OnSA | CS task | Solution to the task |
| e | no | OffSA | CS task | Solution to the task; e.g. trained AI model |
| f | no | OnSA | OffSA | Input data from multiple participants |
| g | no | OffSA | OnSA | A trained AI model to assist participants in problem solving |
| h | no | CS task | Participant | Information about a physical environment or previously acquired experience |

| i | no | External data | OffSA | Data from sources outside the project |
|---|---|---|---|---|
| j | yes | Task selection algorithm | OnSA | Next task or content; might be determined from real time behavioral data (arrow k) |
| k | optional | OnsA | Task selection algorithm | Participants' behavior while solving the task |

**Table 2.** Properties of information transmission between agents from Figure 3.

### The CS task object in the IFD

The diagrams in Fig. 3 show how information is passed between the different agents when solving the *CS task*. The CS task is an abstract object representing both the scientific challenge and the scientists (e.g., the tasks might be dynamically updated). Note that the notion of a task may contain some ambiguity. For instance, in annotation tasks one can either consider "annotation of a single element" or of "the set of elements" as the fundamental "task". Of course, for the researchers the task is the latter, but for the participant it is arguably the former. Thus task selection in the former frame would be called task assistance in the latter frame. Here we define tasks from the perspective of the participants. The task can also represent the physical environment which the participants may be in direct contact with (arrow **h**), such as when the participants take pictures of nature for the iNaturalist project (part of the scientific challenge). The arrow **h** could also represent a previously acquired experience on the matter, for example a general knowledge of shapes required for contributing to the Galaxy Zoo project. The presence and character of the arrow **h** distinguishes between the different *degrees of task digitization* in the individual projects, as discussed previously (see Figure 2). In projects with a high degree of digitization, the information flow through this channel is reduced or even absent, as the CS task can be fully represented by the Online solution algorithm (defined below), simply put, the participants do not need "contact with the physical world" to solve the task.

### The online CS platform in the IFD

In general, the *participants* interact with the CS platform individually (in a private session) via means of an *Online solution algorithm* (OnSA), which is the part of the online CS platform interface where they submit their inputs (arrow **a**) and receive a representation of the task and a feedback on their actions (arrow **b**). Here, algorithm refers to an automated processing of inputs and outputs that may or may not involve AI. This also includes all online (real time) algorithms which may optionally contain AI that help the participants to solve the CS task. For instance, above, we discussed the user-support algorithms in the FoldIt and Quantum Moves 2 games, which both do not contain AI at the current game implementation. The OnSA receives the task either directly (arrow **c**; left diagram: CS task assistance) or via a *Task selection algorithm* (arrow **j**; right diagram: CS task selection) from the CS task.

## The Online-Offline distinction in the IFD

In order to evaluate the HI criteria: *Mutual learning* (see table 1), we need to identify precisely when and where the two stages of Machine Learning (ML) (training and execution; see table 2) take place relative to the participants' input. This is crucial, because it imposes limits on how adaptive the AI can be (e.g., "can it learn during a single session with a CS participant?" or "how does the AI learn from the aggregated knowledge transferred between different participants' sessions?") and limits on how reliable the AI predictions/actions are (too much adaptation/learning may lead to overfitting and bias effects). For this purpose we use two terms: *Online* and *Offline*, defined with respect to the participant's interaction with the algorithmic interface. Here the *online training,* for example, refers to the AI model being updated in real time with data from the individual participants while they engage with the interface (part of OnSA). An Offline AI-model, on the other hand, could be trained on some fixed dataset (acquired via the CS platform or otherwise) resulting in a model that can later interact with participants when executed as a tool in the OnSA or be itself the research outcome (a solution of the CS task). When an offline algorithm is *Assisting in solving the CS task*, we call it an *Offline solution algorithm* (OffSA).

An arrow pointing towards the CS task denotes that a solution was found. The solution might either be directly provided by the OnSA (arrow **d**) or obtained from an OffSA (arrow **e**), where data from multiple participants are loaded from the OnSA (arrow **f**) and aggregated by the OffSA (e.g., the AI model was trained). Alternatively, the AI in the OffSA might be trained on *External data* (arrow **i**; i.e. sources outside of the CS project) and the resulting model might be passed to the OnSA (arrow **g**), where it is executed to assist the participants in solving the task. This is done, for example, in the EyeWire project, where an affinity graph labelling ML algorithm is trained on a scientific grade labeled dataset.

## IFD for CS task and content selection

When an AI is used for *Task selection* (Fig. 3; right panel), it is not part of the OnSA. Rather, it *may* observe the behavior of the individual participants while they are solving the task (arrow **k**). Such AIs use a pre-trained model executed on the real time behavioral data predicting the next suitable content/task (arrow **j**) in order to improve the participants' experience, work efficiency or retention. Alternatively, task selection with AI can be performed offline, forming a common task queue for all participants (disregarding individuality; absent arrow **k**). This approach is adapted, for example, in Galaxy Zoo "Enhanced mode", where several independently pre-trained ML models are executed on the images (tasks), passing the tasks for human classification only in the case of the models' disagreement (low certainty of solution). The task selection algorithm was depicted at the same level as the CS task in Fig. 3 as it can be thought of as an extension of the CS task itself (on the level with the scientists), since it controls the content exposed to the participants and is therefore one of the dynamical components of the scientific challenge.

## Towards an HI hierarchy for CS task assistance

To proceed with our analysis towards HI schemes, we focus on the AI application category *Assisting in solving the CS task* which satisfies the Collectiveness criteria of HI (see table 1) and try to devise a sub-categorisation within it using the HI criterion: *Mutual learning*. Mutual learning requires information being passed bidirectionally between the AI and the human agents (they learn from each other). We therefore need to examine whether the learning and execution stages of the AI are both "in the solution loop" with the CS participants. The answer to that depends crucially on the presence and direction of the information flow between the mandatory OnSA and the optional OffSA, denoted with the two arrows, **f** and **g,** in Figure 3.

Four elementary cases arise from having one, both, or none of the information channels present, as shown in Figure 4. We propose to order these schemes according to how complex they are to implement which goes hand in hand with the degree of mutual learning leading up to HI in Tiers 3 and 4. In other words, these can be described as: 1. AI learning from humans, 2. Humans learning from AI, 3. Mutual learning on a long timescale, and 4. Mutual learning on a short timescale. Here, "humans" are the CS participants only. Our aim here is to discuss the *elementary* schemes that can exist, and in praxis are being combined to form more advanced systems.

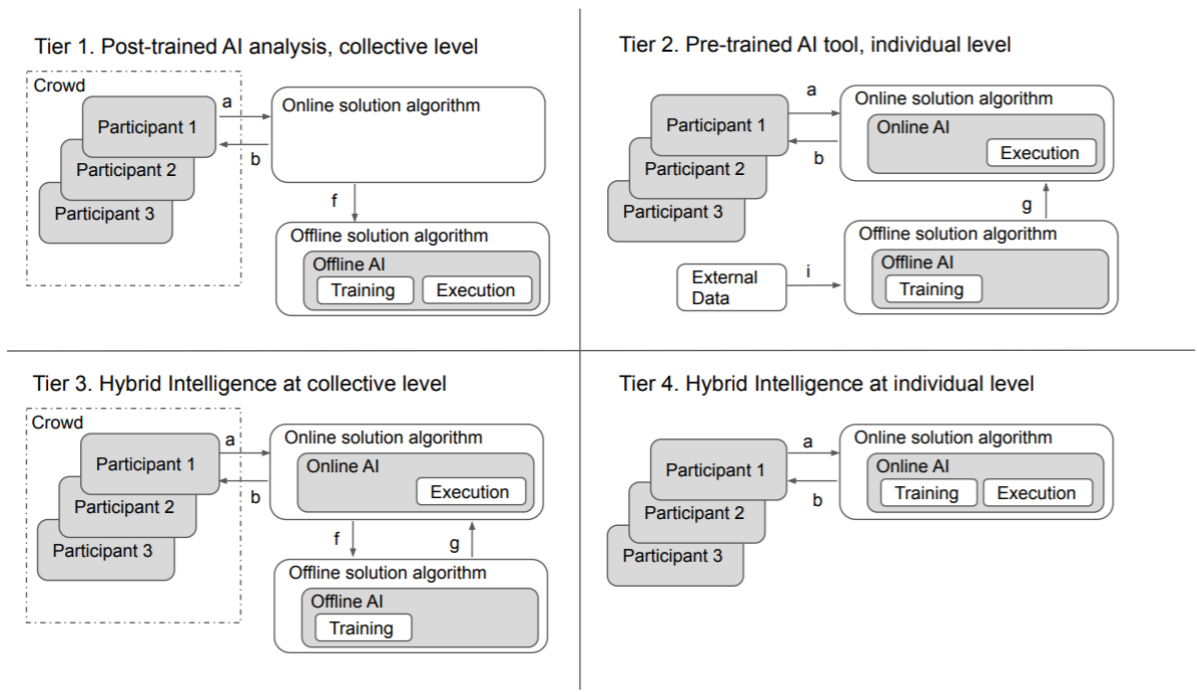

**Figure 4.** Elementary schemes of task assistance by AI, i.e. concrete examples of information flow diagrams based on the generic diagram from Figure 3; left panel. The schemes are organized into tiers leading towards Hybrid Intelligence. They differ primarily by the amount and type of connections between the Online and Offline solution algorithms (arrows **f** and **g**). Arrow labels are consistent with Figure 3. For simplicity, the Citizen Science task object was omitted in these diagrams, though the c, d, e, and h arrows from Figure 3 left would still be relevant here.

*Tier 1: Post-acquisition AI analysis, collective level*

The most common AI task assistance scheme in CS is to train an AI model offline on participant input acquired via an OnSA without AI. This scheme contains only one connection (arrow **f**) and is usually done with data from *multiple participants* acquired over a period of time much longer than the single-task solution time, typically weeks or months, hence the name "collective level", which we visualise with a dash-dotted outline of a *crowd*. The model itself or its execution on unseen data then forms part of the CS task solution. This scheme is very convenient from the ML perspective, because the quantity and quality of participant data is often not known in advance. AI model architecture can be chosen accordingly (post-factum) in order to prevent *overfitting* (attributing meaning to noise in the data). This scheme is applied for example in Stall Catchers (*Stall Catchers 2020*), Galaxy Zoo (Lintott et al. 2008), EyeWire (Kim et al. 2014), Quantum Moves 2 (Jensen et al. 2021), Eterna (Andreasson et al. 2019), Phylo (Kawrykow et al. 2012), and Fraxinus (Rallapalli et al. 2015).

*Tier 2: Pre-trained AI tool, individual level*

In this tier, a fixed, pre-trained AI model is executed as a component of the OnSA, assisting the participants with the task. In other words, participants can learn from the AI, but the AI cannot learn from the participants. The model was previously trained in an OffSA using obtained outside of the CS platform and passed into the OnSA (arrow **g**). This scheme can increase efficiency and quality of the task solutions, combining strengths of humans and machines to provide superior outcomes (HI criteria of *Collectiveness* and *Superiority*). Among the investigated projects, this scheme is applied in EyeWire (Kim et al. 2014) and iNaturalist (*iNaturalist* 2021). It is worth mentioning in Foldit (Cooper et al. 2010) and Quantum Moves (Jensen et al. 2021) participants can engage with algorithmic optimizers (not AI) leading to a similar increase in participants' learning and productivity.

*Tier 3: Hybrid Intelligence at collective level*

In this tier, the CS system satisfies all three criteria of HI, including mutual learning of the heterogeneous agents as well as learning of the system as a whole. Similarly to Tier 2, the OnSA contains a fixed AI model which assists the participants in solving the task; however, the AI model is trained on the participant data acquired earlier on the very same platform (arrow **f**). The information flow through the OffSA therefore forms a closed loop as the re-trained/updated AI model is fed back to the OnSA (arrow **g**). While the AI does not learn in real time while the individual participants interact with the OnSA it can learn over weeks or months at the collective level, as the model is periodically updated with batches of new data from multiple participants. On an abstract level, the crowd plays the role of the "human agent", since the AI is learning from the crowd and the crowd experiences that the AI improves over time. Individuals from the crowd might experience learning of the AI as well if they stay engaged on the CS platform for multiple training cycles. To our knowledge, the only project adopting HI of this type is iNaturalist (*iNaturalist*, 2021), where participants can use an AI model to classify their images and the model is periodically updated with human-labeled data from the platform. Such an implementation is also underway in the Stall Catchers project and is described in the outlook.

*Tier 4: Hybrid Intelligence at individual level*

Tier 4 also satisfies all three components of HI. However, the OnSA contains an AI model that is trained and executed while individual participants interact with it, allowing for mutual learning of the human and AI agents in real time. The basic form of this scheme does not contain an OffSA, implying the absence of arrows **f** and **g**. To our knowledge, no CS project uses this type of scheme. Although, judging by its information flow diagram, it appears particularly simple, the presence of both training and execution in real time interaction with participants raises many practical and technical issues. In order for the participants to benefit beyond what is gained in Tier 3, they must be able to experience the effect of their actions on the AI, otherwise it could be trained offline. This requires a very high learning rate for the AI, raising the risk of overfitting and ceasing to be useful on the task (i.e. the AI agreeing with the individual on all actions).

Due to large numbers of participants in CS, the effect of an individual on the final result, e.g. an AI model, tends to be rather small (often intentionally so). Nevertheless, a quickly learning AI might still be useful in some tasks, for example when it is crucial to extract the strategy of an individual. One could think of it as a supervised automation of the task solution. For example in Decodoku, the participants are asked to describe their strategies verbally. With the appropriate AI assistance, the meta cognitive reflection could potentially be enhanced. In general, projects with high digitization containing a closed model of the task are well suited for *Tier 4. Hybrid Intelligence at individual level*. As seen in Figure 2, high digitization tasks addressed here are in principle solvable with RL (an optimization problem). In one such scenario, an AI might propose the next actions to take, while humans either accept the proposal, modify it or choose a different action altogether. The AI would therefore be learning strategies from an individual with an immediate feedback loop, or in the ML terminology, the participants would provide an adaptable reward function for an RL algorithm. This could provide better outcomes than just training a model on many "games" from a certain player, the AI can also influence the player by for example increasing consistency of the players actions or on the contrary, sparking new ideas. In addition, having the AI train online with humans in the loop could reduce the issue of *perverse instantiation* which gives rise to bizarre, unwanted solutions in complex problems with an a priori defined reward function (Bostrom 2014).

Although these Tiers of AI task assistance tend to be progressively more challenging to implement, it is not granted that implementing a higher tier is always worth the extra effort for research outcomes or is feasible and beneficial in every CS scenario. The chosen AI assistance scheme should be carefully considered, taking into account the character and amount of data as well as the AI methodology applicable to the task. For example in Supervised Learning tasks such as in iNaturalist (iNaturalist 2021), Galaxy Zoo (Lintott et al. 2008), Stall Catchers (Nugent 2019) and EyeWire (Kim et al. 2014), strong individual participant effects on the model are not desirable, as this can introduce large decision biases. Similarly, presence of certain types of AI agents in the OnSA may have a detrimental effect on the outcomes, e.g. biasing participants to solve the task in a specific way rather than contributing creatively to the project. Implementing higher tiers of AI task assistance therefore always bears extra overhead on design considerations, AI architecture choice, and ultimately maintenance of the system.

## 5. Conclusion and Outlook

Herein we have described a value proposition for HI in CS. In acknowledging uncertainties about HI architectures and potential outcomes, we introduce a rich space of research opportunities that could help realize the tremendous potential of adaptive and synergistic human+AI relationships.

A key defining quality of HI is the mutual learning that exists among the AI and human components of the system. Such integration not only allows for, but necessitates the co-evolution of individual components (AI and human alike) with each other and the systems to which they contribute. Thus, we are entering into uncharted AI territory, where our best hope for advancing the field may require bootstrapping. In other words, the potential complexity of these systems suggests an opportunity to use HI itself to improve our understanding of HI. Asking why a given CS project is not solved entirely algorithmically can be a path to identifying new modes of human-machine problem solving by discovering suitable existing machine technologies and as well as to deeper appreciation of the distinctly human contribution in areas where current machine technology falls short.

Areas of great potential for human augmentation are applications of RL to optimization challenges, SL to classification tasks, and smart sensors to participatory sensing tasks. On the other hand, tasks in which full automation is beyond current technological reach typically tap into common sense, hierarchical thinking or meta-cognitive reflection and full human-level mobility combined with environmental sensing and domain knowledge.

How then, as a community of scientists, data science practitioners, and domain researchers are we to pursue these questions concertedly and efficiently? How, for example, can we study various information flow architectures such as those depicted in Figure 4 without having to manually design and code each one and then recruit human participants? A potential means of rapid experimentation could be through an online research platform to enable rapid development and testing of various candidate HI architectures for any CS project. Currently the developers of Stall Catchers (Bracko, et al. 2020, Falkenhain et al. 2020) are developing a platform to become the basis for a modular project builder that integrates with new human/AI research capabilities. This resulted in an experimentation dashboard that makes it possible to clone an existing CS project into a "sandbox" version, design a human/AI study, select data, invite participants, and run experiments from a single interface and without writing a single line of code. This research platform is part of a broader initiative called "Civium" (Borfitz 2019, Michelucci 2019, Vepřek et al. 2020), which seeks to make all advanced information processing systems, including hybrid intelligence research and applications, more transparent, trustworthy, and sustainable.

# 6. Supplemental Files List

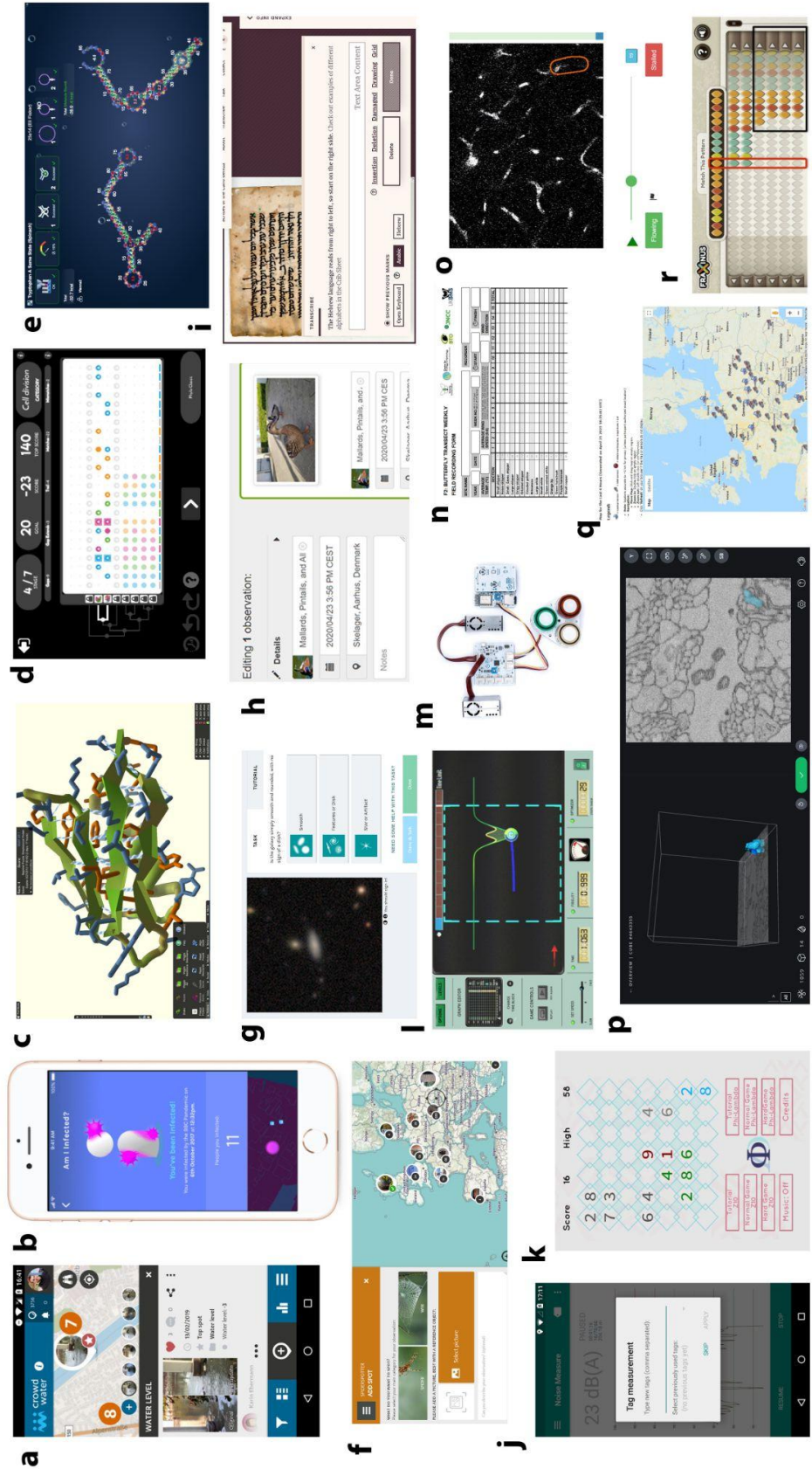

**Supplement 1**: Screenshots of the projects. a. Crowd Water (Crowd Water n.d.) . b. BBC Pandemic (Manson, E. n.d.). c. Foldit (Foldit n.d.). d. Phylo (Phylo 2021). e. Eterna (Eterna 2018). f. Spider Spotter (Spider Spotter 2021). g. Galaxy Zoo (Galaxy Zoo 2021). h. iNaturalist (iNaturalist 2021). i. Scribes of the Cairo Geniza (Scribes of the Cairo Geniza 2021). j. Noisetube (Noisetube 2021). k. Decodoku (Decodoku 2021). l. Quantum Moves 2 (Ahmed n.d.). m. Smart Citizen (Camprodon et al. 2019. n. UK Butterfly Monitoring Scheme UKBMS (n.d.). o. Stall Catchers (Stall Catchers 2021). p. Eyewire (Eyewire 2021). q. Quake Catcher (Quake Catcher n.d.). r. Fraxinus (Rallapalli 2015.

| Name | General project description | Core participant task | Web reference |
|---|---|---|---|
| *Annotation tasks* | | | |
| Galaxy Zoo | Classification of images of galaxies based on an evolving scheme devised by scientists who use the resulting classifications as part of their studies (Ponti et al., 2018). | Classifying images of galaxies based on particular features in a decision tree | https://www.zooniverse.org/projects/zookeeper/galaxy-zoo/ <br> Ponti, M., Hillman, T., Kullenberg, C., & Kasperowski, D. (2018). Getting it Right or Being Top Rank: Games in Citizen Science. *Citizen Science: Theory and Practice*, *3*(1), 1. https://doi.org/10.5334/cstp.101 <br> *Zooniverse*. (n.d.). Retrieved August 21, 2020, from https://www.zooniverse.org/projects/zookeeper/galaxy-zoo/ |
| Eyewire | Mapping the 3D structure of neurons in the brain and reconstructing neural circuits (*Explore \| EyeWire*, n.d.) from serial electron microscope images ("*Eyewire*," 2020) to discover how neurons connect and network to process information. | Solving 2D and 3D puzzles: Identifying and coloring axons of neurons in 2D view of 3D cubes using microscope images and pattern recognition skills | https://eyewire.org/explore <br> *Explore \| EyeWire*. (n.d.). Retrieved August 21, 2020, from https://eyewire.org/explore <br> *Eyewire*. (2020). In *Wikipedia*. https://en.wikipedia.org/w/index.php?title=Eyewire&oldid=950539218 |
| Stall Catchers | Part of the EyesOnAlz project, an online game that helps to speed up Alzheimer's disease research by making it available to anyone to analyze data of microscopic recordings of vessels in the brains of mice inserted with the human gene of the disease (*Join a Global Game That's Trying to Cure Alzheimer's*, n.d.). | Looking at movies clips from the brains of mice and trying to identify blood vessels as flowing or stalled (clogged) | https://stallcatchers.com/main <br> *Join a global game that's trying to cure Alzheimer's*. (n.d.). Stall Catchers. Retrieved August 21, 2020, from https://stallcatchers.com |
| *Optimization tasks* | | | |
| Quantum Moves | Finding the optimal solution for Quantum Mechanics evolution of wavefunction in a dynamical potential in the shortest possible time/duration. (Jensen et al., 2020, p.4). | Transferring atoms the best possible way from a specified initial state to the desired target state within very short timescales | https://www.scienceathome.org/games/quantum-moves-2/ <br> Jensen, J. H. M., Gajdacz, M., Ahmed, S. Z., Czarkowski, J. H., Weidner, C., Rafner, J., Sørensen, J. J., Mølmer, K., & Sherson, J. F. (2020). Crowdsourcing human common sense for quantum control. *ArXiv:2004.03296 [Quant-Ph]*. http://arxiv.org/abs/2004.03296 <br> Games, S. : C. S. (n.d.). *ScienceAtHome \| Games \| |

| | | | *Quantum Moves 2*. ScienceAtHome.Org. Retrieved August 21, 2020, from https://www.scienceathome.org/games/quantum-moves-2/ |
|---|---|---|---|
| EteRNA | Capitalizing on the collective intelligence of EteRNA players to answer fundamental questions about RNA folding mechanics (*EteRNA | Crowdsourcing New RNA Designs | CitizenScience.Gov*, n.d.): Understanding and mastering the synthesis of RNA molecules and the RNA conformation for multiple medical, therapeutic and biotechnological applications (Lafourcade et al., 2015, p.5). | 2D puzzle solving game with the four bases of RNA: designing elaborate structures, including knots, lattices and switches (*EteRNA | Crowdsourcing New RNA Designs | CitizenScience.Gov*, n.d.) | https://eternagame.org/home/<br>*EteRNA | Crowdsourcing New RNA Designs | CitizenScience.gov*. (n.d.). Retrieved August 21, 2020, from https://www.citizenscience.gov/eterna/<br>Lafourcade, M., Joubert, A., & Le Brun, N. (2015). *Games with a Purpose (GWAPS)* (1st ed.). John Wiley & Sons, Ltd. https://doi.org/10.1002/9781119136309<br>*Eterna*. (n.d.). Retrieved August 21, 2020, from https://eternagame.org/ |
| FoldIt | Crowdsource problems in protein modelling: Creating predictive models of three-dimensional structures of proteins from their amino acid composition (Lafourcade et al., 2015, p.2) to understand how a mutation occurs at the level of the spatial conformation, and develop appropriate therapies (Lafourcade et al., 2015, p.2). | Protein folding puzzle game: players are presented with an unstructured amino acid sequence and challenged to determine its native conformation (Koepnick et al., 2019, p.390) | https://fold.it/<br>Lafourcade, M., Joubert, A., & Le Brun, N. (2015). *Games with a Purpose (GWAPS)* (1st ed.). John Wiley & Sons, Ltd. https://doi.org/10.1002/9781119136309<br>Koepnick, B., Flatten, J., Husain, T., Ford, A., Silva, D.-A., Bick, M. J., Bauer, A., Liu, G., Ishida, Y., Boykov, A., Estep, R. D., Kleinfelter, S., Nørgård-Solano, T., Wei, L., Players, F., Montelione, G. T., DiMaio, F., Popović, Z., Khatib, F., … Baker, D. (2019). De novo protein design by citizen scientists. *Nature*, *570*(7761), 390–394. https://doi.org/10.1038/s41586-019-1274-4<br>*Solve Puzzles for Science | Foldit*. (n.d.). Retrieved August 21, 2020, from https://fold.it/ |
| Decodoku | 2-D puzzle solving game simulating the build-up of unwanted quantum interactions in quantum computers to learn about the cognitive strategies and heuristics players use to correct these errors, which can later be used for improving algorithms used in quantum error correction (*Scientists Need You to Join the Phi-Lambda Mission and Make Quantum Computers Work*, n.d.). | Adding up the numbers in the grids, where multiples of 10 disappear and new errors may arise until the game is over, then the players can provide an explanation on the thought processes and strategies they used during playing. | https://decodoku.com/<br>*Scientists need you to join the Phi-Lambda mission and make quantum computers work*. (n.d.). Retrieved August 21, 2020, from https://www.zmescience.com/science/physics/phi-lambda-mission/<br>*Decodoku*. (n.d.). Retrieved August 21, 2020, from https://decodoku.com/ |
| Fraxinus | Fighting the Ash dieback disease, a disease of ash trees, by identify regions of DNA sequences that show characteristics like resistance, which might then be bred into a new disease-resistant variety. (Tsouvalis, 2015, p.2). | A pattern-matching Facebook game: players aim to find the best match between color patterns | https://teamcooper.co.uk/work/fraxinus/<br>Tsouvalis, J. (2015). *How social and citizen science help challenge the limits of the biosecurity approach: The case of ash dieback*.<br>*Team Cooper » Fraxinus*. (n.d.). Retrieved August 21, 2020, from https://teamcooper.co.uk/work/fraxi |

| | | | nus/ |
|---|---|---|---|
| Phylo | A game about multiple sequence alignment optimization: Players solve pattern-matching puzzles that represent nucleotide sequences of different phylogenetic taxa to optimize alignments over a computer algorithm ("Play Phylo, Solve DNA Puzzle and Help Genetic Disease Research," n.d.). | Solving pattern-matching puzzles of colored blocks: maximizing color matches across columns for best vertical alignments while minimizing gaps within sequences. | https://phylo.cs.mcgill.ca/<br>Play Phylo, solve DNA puzzle and help genetic disease research. (n.d.). *Citizen Science Games*. Retrieved August 23, 2020, from https://citizensciencegames.com/games/phylo/<br>*Phylo DNA Puzzle*. (n.d.). Retrieved August 23, 2020, from https://phylo.cs.mcgill.ca/ |
| | | *Physical tasks* | |
| iNaturalist | A community of scientists, naturalists and citizen scientists and tools to create research quality data for scientists working to better understand and protect nature. (*About · iNaturalist*, n.d.) | Recording and sharing observations of the nature using an app | https://www.inaturalist.org/<br>*About · iNaturalist*. (n.d.). iNaturalist. Retrieved August 23, 2020, from https://www.inaturalist.org/pages/about<br>*INaturalist*. (n.d.). iNaturalist. Retrieved August 23, 2020, from https://www.inaturalist.org/ |
| Smart Citizen | A kit to collect data and a platform to connect people to collectively address and find solutions to local environmental problems (*Smart Citizen*, n.d.). | Recording real-time environmental data and share it with the community | https://digitalsocial.eu/case-study/9/smart-citizen<br>*Smart Citizen*. (n.d.). Retrieved August 23, 2020, from https://digitalsocial.eu/case-study/9/smart-citizen |
| UK Butterfly Monitoring Scheme | Monitoring changes in the abundance of butterflies throughout the UK based on a well-established and enjoyable method to understand trends in insect populations and answer policy questions relating to status and trends in biodiversity. (*United Kingdom Butterfly Monitoring Scheme | Home Page*, n.d.) | Selecting site, designing route and recording and submitting data weekly (Pollard walks) | https://www.ukbms.org/<br>*United Kingdom Butterfly Monitoring Scheme | Home Page*. (n.d.). Retrieved August 23, 2020, from https://www.ukbms.org/ |
| NoiseTube | Monitoring noise pollution to inform the community, create collective, city-wide noise maps, improve policy making with regards to noise level, and research soundscape perception. (*NoiseTube*, n.d.) | Monitoring noise level in surroundings using an app and tagging the measurements (e.g. subjective level of annoyance, source of sound) | http://www.noisetube.net/index.html#&panel1-1<br>*NoiseTube*. (n.d.). Retrieved August 23, 2020, from http://www.noisetube.net/index.html#&panel1-1 |
| Spider Spotter | Studying spider evolution in the city in real time how spiders adapt to the increased heat and other special circumstances in the city to inform research on climate change and potentially discover new ways to adapt to the changing environment (*Info - SpiderSpotter*, n.d.). | Taking pictures of spiders and/or their web with a reference object using a smartphone app and/or analyse photos on the website and calculate the color and length of the spider or web (*Info - SpiderSpotter*, n.d.). | https://www.spiderspotter.com/en/<br>*Home—SpiderSpotter*. (n.d.). Retrieved August 28, 2020, from https://www.spinnenspotter.be/en/<br>*Info—SpiderSpotter*. (n.d.). Retrieved August 28, 2020, from https://www.spiderspotter.com/en/info/spin-city |

**Supplement 2**. Projects, descriptions, and sources.

## 7. Competing Interests

The author(s) has/have no competing interests to declare.

## 8. References

Ahmed, SZ. (n.d.)
    https://www.scienceathome.org/games/quantum-moves-2/about-quantum-moves-2/

Akata, Z., Balliet, D., de Rijke, M., Dignum, F., Dignum, V., Eiben, G., Fokkens, A., Grossi, D., Hindriks, K., Hoos, H., Hung, H., Jonker, C., Monz, C., Neerincx, M., Oliehoek, F., Prakken, H., Schlobach, S., Gaag, L. van der, Harmelen, F. van, … Welling, M. 2020. A Research Agenda for Hybrid Intelligence: Augmenting Human Intellect With Collaborative, Adaptive, Responsible, and Explainable Artificial Intelligence. *Computer*, 53(8): 18–28. DOI: https://doi.org/10.1109/MC.2020.2996587

Anderson-Lee, J., Fisker, E., Kosaraju, V., Wu, M., Kong, J., Lee, J., Lee, M., Zada, M., Treuille, A., & Das, R. 2016. Principles for Predicting RNA Secondary Structure Design Difficulty. Journal of Molecular Biology. *Journal of molecular biology*, 428(5), 748-757. DOI: https://doi.org/10.1016/j.jmb.2015.11.013

Andreasson, JOL., Gotrik, MR., Wu, MJ., Wayment-Steele, HK., Kladwang, W., Portela, F., Wellington-Oguri, R., Participants, E., Das, R., Greenleaf, W.J. 2019. Crowdsourced RNA design discovers diverse, reversible, efficient, self-contained molecular sensors (preprint). *Synthetic Biology*. DOI: https://doi.org/10.1101/2019.12.16.877183

Auernhammer, J. 2020. Design Research in Innovation Management: a pragmatic and human-centered approach. *R&D Management*, 50(3): 412-428.

Baltz, EA., Trask, E., Binderbauer, M., Dikovsky, M., Gota, H., Mendoza, R., Platt, JC., Riley, PF., 2017. Achievement of Sustained Net Plasma Heating in a Fusion Experiment with the Optometrist Algorithm. *Scientific reports*, 7(1): 1-7. DOI: https://doi.org/10.1038/s41598-017-06645-7

Bellman, R. 1978. *An Introduction to Artificial Intelligence: Can Computers Think?* Boyd & Fraser Publishing Company.

Benedikt, L., Joshi, C., Nolan, L., Henstra-Hill, R., Shaw, L., Hook, S., 2020. Human-in-the-loop AI in government: a case study. In *Proceedings of the 25th International Conference on Intelligent User Interfaces*. Cagliari Italy in March 2020, pp. 488–497.

Bohrdt, A., Chiu, C.S., Ji, G., Xu, M., Greif, D., Greiner, M., Demler, E., Grusdt, F., Knap, M. 2019. Classifying snapshots of the doped Hubbard model with machine learning. *Nature Physics,* 15(9): 921–924. DOI: https://doi.org/10.1038/s41567-019-0565-x

Bonabeau, E.. Dorigo, M. & Theraulaz, G. 2001. Swarm Intelligence : From Natural to Artificial Systems. Oxford, New York: Oxford University Press.

Bonney, R., Ballard, H., Jordan, R., McCallie, E., Phillips, T., Shirk, J., Wilderman, CC., Russell, P. (1995). The Global Brain Awakens: Our next evolutionary leap (Global Brain). Inc., CA.

Bonney, R., Cooper, C.B., Dickinson, J., Kelling, S., Phillips, T., Rosenberg, KV., Shirk, J., 2009a.

Public Participation in Scientific Research: Defining the Field and Assessing Its Potential for Informal Science Education. A CAISE Inquiry Group Report.

Bonney, R., Cooper, C.B., Dickinson, J., Kelling, S., Phillips, T., Rosenberg, KV., Shirk, J., 2009b. Citizen Science: A Developing Tool for Expanding Science Knowledge and Scientific Literacy. *BioScience*, 59(11): 977–984. DOI: https://doi.org/10.1525/bio.2009.59.11.9

Bonney, R., Phillips, TB., Ballard, HL., Enck, JW. 2016. Can citizen science enhance public understanding of science? *Public Understanding of Science,* 25(1): 2–16. DOI: https://doi.org/10.1177/0963662515607406

Borfitz, D. 2019. A New Marketplace For Machine Learning Researchers—And Citizen Scientists. *Bio IT World*. Available at https://www.bio-itworld.com/news/2019/10/22/a-new-marketplace-for-machine-learning-researchers-and-citizen-scientists (Last accessed 21 April 2021)

Bostrom, N., 2014. Superintelligence: Paths, dangers, strategies. Oxford, New York: Oxford University Press.

Bracko, O., Vinarcsik, LK., Hernández, JCC., Ruiz-Uribe, NE., Haft-Javaherian, M., Falkenhain, K., Ramanauskaite, EM., Ali, M., Mohapatra, A., Swallow, MA., Njiru, BN., Muse, V., Michelucci, PE., Nishimura, N., Schaffer, CB. 2020. High fat diet worsens Alzheimer's disease-related behavioral abnormalities and neuropathology in APP/PS1 mice, but not by synergistically decreasing cerebral blood flow. *Scientific Reports,* 10(1): 1–16. DOI: https://doi.org/10.1038/s41598-020-65908-y

Camprodon, G., González, Ó., Barberán, V., Pérez, M., Smári, V., de Heras, MÁ., Bizzotto, A. 2019. Smart Citizen Kit and Station: An open environmental monitoring system for citizen participation and scientific experimentation. *HardwareX*, 6: e00070. DOI: https://doi.org/10.1016/j.ohx.2019.e00070

Causer, T., & Terras, M. 2014. 'Many Hands Make Light Work. Many Hands Together Make Merry Work'1: Transcribe Bentham and Crowdsourcing Manuscript Collections. In Ridge, M. (eds.) *Crowdsourcing our Cultural Heritage*. Farnham:Ashgate. 57-88.

Ceccaroni, L., Bibby, J., Roger, E., Flemons, P., Michael, K., Fagan, L., Oliver, J.L. 2019. Opportunities and Risks for Citizen Science in the Age of Artificial Intelligence. *Citizen Science: Theory and Practice,* 4(1): 29. DOI: https://doi.org/10.5334/cstp.241

Chen, NC., Drouhard, M., Kocielnik, R., Suh, J., Aragon, CR. 2018. Using Machine Learning to Support Qualitative Coding in Social Science: Shifting the Focus to Ambiguity. *ACM Transactions on Interactive Intelligent Systems*. 8(2): 1-20. DOI: https://doi.org/10.1145/3185515

Christiano, PF., Leike, J., Brown, T., Martic, M., Legg, S., & Amodei, D. 2017. Deep Reinforcement Learning from Human Preferences. In Guyon, I., Luxburg, UV., Bengio, S. Wallach, H. Fergus, R. Vishwanathan, S. & Garnett, R. (Eds.) *Advances in Neural Information Processing Systems 30*, Montreal: Curran Associates Inc. pp. 4299–4307. http://papers.nips.cc/paper/7017-deep-reinforcement-learning-from-human-preferences.pdf

Cochran, ES., Lawrence, J.F., Christensen, C., Jakka, RS. 2009. The Quake-Catcher Network: Citizen Science Expanding Seismic Horizons. *Seismological Research Letters*, 80(1): 26–30.

DOI: https://doi.org/10.1785/gssrl.80.1.26

Cooper, S., Khatib, F., Treuille, A., Barbero, J., Lee, J., Beenen, M., Leaver-Fay, A., Baker, D., Popović, Z., Players, F. 2010. Predicting protein structures with a multiplayer online game. *Nature*, 466: 756–760. DOI: https://doi.org/10.1038/nature09304

Copeland, BJ. 1997. The Broad Conception of Computation. *American Behavioral Scientist*, 40(6): 690–716. DOI: https://doi.org/10.1177/0002764297040006003

Correia, A., Paredes, H., Schneider, D., Jameel, S., & Fonseca, B. 2019. Towards Hybrid Crowd-AI Centered Systems: Developing an Integrated Framework from an Empirical Perspective. In *2019 IEEE International Conference on Systems, Man and Cybernetics (SMC)*, Bari, Italy on 6-9 October 2019, pp. 4013–4018. DOI: https://doi.org/10.1109/SMC.2019.8914075

Crowd Water. (n.d.) .Available at
https://crowdwater.ch/en/crowdwaterapp-en/#NeuerSpotFliessgew%C3%A4ssertyp

Crowston, K., Østerlund, C., Lee, TK., Jackson, C., Harandi, M., Allen, S., Bahaadini, S., Coughlin, S., Katsaggelos, A. K., Larson, S. L., Rohani, N., Smith, J. R., Trouille, L., & Zevin, M. 2020. Knowledge Tracing to Model Learning in Online Citizen Science Projects. *IEEE Transactions on Learning Technologies*, 13(1): 123–134. DOI: https://doi.org/10.1109/TLT.2019.2936480

Curtis, V. 2015. Motivation to Participate in an Online Citizen Science Game: A Study of Foldit. *Science Communication*, 37(6): 723–746. DOI:
https://doi.org/10.1177/1075547015609322

D'Hondt, E., Stevens, M., Jacobs, A. 2013. Participatory noise mapping works! An evaluation of participatory sensing as an alternative to standard techniques for environmental monitoring. Pervasive and Mobile Computing, *Special issue on Pervasive Urban Applications,* 9(5): 681–694. DOI: https://doi.org/10.1016/j.pmcj.2012.09.002

Dalgaard, M., Motzoi, F., Sørensen, J. J., & Sherson, J. 2020. Global optimization of quantum dynamics with AlphaZero deep exploration. *Npj Quantum Information*, 6(1), 1–9. DOI: https://doi.org/10.1038/s41534-019-0241-0

Danielsen, F., Burgess, ND., Coronado, I, Enghoff, M, Holt, S, Jensen, PM., Poulsen, MK., & Rueda, RM. 2018. The value of indigenous and local knowledge as citizen science. In Hecker, S., Haklay, M., Bowser, A., Makuch, Z., & Vogel, J. (eds.) *Citizen Science: Innovation in Open Science*, *Society and Policy*. London: UCL Press. 110–123. DOI: https://doi.org/10.14324/111.9781787352339

Danielsen, F., Topp-Jørgensen, E., Levermann, N., Løvstrøm, P., Schiøtz, M., Enghoff, M., & Jakobsen, P. 2014. Counting what counts: Using local knowledge to improve Arctic resource management. *Polar Geography*, 37(1): 69–91. DOI:
https://doi.org/10.1080/1088937X.2014.890960

Das, R., Keep, B., Washington, P., & Riedel-Kruse, IH. 2019. Scientific Discovery Games for Biomedical Research. *Annual Review of Biomedical Data Science*, 2(1): 253–279. DOI: https://doi.org/10.1146/annurev-biodatasci-072018-021139

Decodoku. 2021. Available at https://decodoku.itch.io/decodoku

Dellermann, D., Ebel, P., Söllner, M., & Leimeister, JM. 2019. Hybrid Intelligence. *Business & Information Systems Engineering*, 61(5): 637–643. DOI:

https://doi.org/10.1007/s12599-019-00595-2

Dennis, EB., Morgan, BJT., Brereton, TM., Roy, DB., Fox, R. 2017. Using citizen science butterfly counts to predict species population trends. *Conservation Biology*, 31(6), 1350–1361. DOI: https://doi.org/10.1111/cobi.12956

Eager, J., Whittle, M., Smit, J., Cacciaguerra, G., Lale-Demoz, E., 2020. Opportunities of Artificial Intelligence. PE 652.713. European Union.
http://www.europarl.europa.eu/supporting-analyses

Eiben, CB., Siegel, JB., Bale, JB., Cooper, S., Khatib, F., Shen, BW., Players, F., Stoddard, BL., Popovic, Z., Baker, D. 2012. Increased Diels-Alderase activity through backbone remodeling guided by Foldit players. *Nature Biotechnology,* 30(2): 190–192. DOI: https://doi.org/10.1038/nbt.2109

Eterna. 2018. Available at https://eternagame.org/news/8997813?sort=blog

Eyewire. 2021. Available at https://eyewire.org/

Falkenhain, K., Ruiz-Uribe, N. E., Haft-Javaherian, M., Ali, M., Stall Catchers, Michelucci, PE., … & Bracko, O. 2020. A pilot study investigating the effects of voluntary exercise on capillary stalling and cerebral blood flow in the APP/PS1 mouse model of Alzheimer's disease. *PloS one*, 15(8): e0235691. DOI: 10.1371/journal.pone.0235691

Foldit. (n.d.). Available at https://fold.it/portal/info/about

Fortson, L., Masters, K., Nichol, R., Borne, K., Edmondson, E., Lintott, C., Raddick, J., Schawinski, K., & Wallin, J. 2011. Galaxy Zoo: Morphological Classification and Citizen Science. *Advances in Machine Learning and Data Mining for Astronomy*. 2012: 213-236. arXiv:1104.5513

Galaxy Zoo. 2021. Available at https://www.zooniverse.org/projects/zookeeper/galaxy-zoo/classify

Godines, D., Bachelet, E., Narayan, G., Street, R.A. 2019. A machine learning classifier for microlensing in wide-field surveys. *Astronomy and Computing*, 28: 100298. DOI: https://doi.org/10.1016/j.ascom.2019.100298

Haklay, M. 2013. Citizen Science and Volunteered Geographic Information: Overview and Typology of Participation, In Sui, D., Elwood, S., Goodchild, M. (Eds.) *Crowdsourcing Geographic Knowledge: Volunteered Geographic Information (VGI) in Theory and Practice.* Dordrecht, Netherlands:Springer, pp. 105–122. DOI: https://doi.org/10.1007/978-94-007-4587-2_7

Heaven, D., 2019. Why deep-learning AIs are so easy to fool. *Nature*, 574(7777): 163–166. https://doi.org/10.1038/d41586-019-03013-5

Heck, R., Vuculescu, O., Sørensen, J.J., Zoller, J., Andreasen, M.G., Bason, M.G., Ejlertsen, P., Elíasson, O., Haikka, P., Laustsen, J.S., Nielsen, L.L., Mao, A., Müller, R., Napolitano, M., Pedersen, M.K., Thorsen, A.R., Bergenholtz, C., Calarco, T., Montangero, S., Sherson, J.F., 2018. Remote optimization of an ultracold atoms experiment by experts and citizen scientists. *Proceedings of the National Academy of Sciences,* 115: E11231–E11237. DOI: https://doi.org/10.1073/pnas.1716869115

Hindman, M. 2015. Building Better Models: Prediction, Replication, and Machine Learning in the Social Sciences. *The ANNALS of the American Academy of Political and Social Science*, 659(1): 48–62. DOI: https://doi.org/10.1177/0002716215570279

Holzinger, A. 2016. Interactive machine learning for health informatics: when do we need the human-in-the-loop? *Brain Informatics*, 3(2): 119–131. DOI: 10.1007/s40708-016-0042-6 https://ukbms.org/sites/default/files/downloads/UKBMS%20F2%20Weekly%20recording%20form.pdf

Huang, J., Hmelo-Silver, CE., Jordan, R., Gray, S., Frensley, T., Newman, G., Stern, MJ. 2018. Scientific discourse of citizen scientists: Models as a boundary object for collaborative problem solving. *Computers in Human Behavior*, 87: 480–492. DOI: https://doi.org/10.1016/j.chb.2018.04.004

iNaturalist, 2021. iNaturalist. Available at: https://www.inaturalist.org/ (Last accessed 21 April 2021)

Iten, R., Metger, T., Wilming, H., del Rio, L., Renner, R., 2020. Discovering Physical Concepts with Neural Networks. *Physical Review Letters*, 124(1): 010508. DOI: https://doi.org/10.1103/PhysRevLett.124.010508

Jensen, JHM., Gajdacz, M., Ahmed, SZ., Czarkowski, JH., Weidner, C., Rafner, J., Sørensen, JJ., Mølmer, K., Sherson, JF., 2021. Crowdsourcing human common sense for quantum control. Physical Review Research, 3(1): 013057. DOI: https://doi.org/10.1103/PhysRevResearch.3.013057

Kahneman, D. 2011. *Thinking, fast and slow.* New York: Farrar, Straus and Giroux.

Kamar, E. 2016. Directions in Hybrid Intelligence: Complementing AI Systems with Human Intelligence. In Proceedings of the Twenty-Fifth International Joint Conference on Artificial Intelligence, New York, New York, USA on 9–15 July 2016, pp. 4070–4073. URL: https://www.microsoft.com/en-us/research/publication/directions-hybrid-intelligence-complementing-ai-systems-human-intelligence/

Kamar, E., Manikonda, L. 2017. Complementing the Execution of AI Systems with Human Computation. *In AAAI Workshops*. San Francisco, California, USA on 4-5 February 4 2017. URL: http://aaai.org/ocs/index.php/WS/AAAIW17/paper/view/15092

Kawrykow, A., Roumanis, G., Kam, A., Kwak, D., Leung, C., Wu, C., Zarour, E., Players, P., Sarmenta, L., Blanchette, M., Waldispühl, J. 2012. Phylo: A Citizen Science Approach for Improving Multiple Sequence Alignment. *PLOS ONE*, 7(3): e31362. DOI: https://doi.org/10.1371/journal.pone.0031362

Khatib, F., DiMaio, F., Cooper, S., Kazmierczyk, M., Gilski, M., Krzywda, S., Zabranska, H., Pichova, I., Thompson, J., Popović, Z., Jaskolski, M., Baker, D. 2011. Crystal structure of a monomeric retroviral protease solved by protein folding game players. *Nature Structural & Molecular Biology*, 18(10): 1175–1177. DOI: https://doi.org/10.1038/nsmb.2119

Kim, JS., Greene, MJ., Zlateski, A., Lee, K., Richardson, M., Turaga, SC., Purcaro, M., Balkam, M., Robinson, A., Behabadi, B.F., Campos, M., Denk, W., Seung, H.S. 2014. Space–time wiring specificity supports direction selectivity in the retina. *Nature* 509(7500): 331–336. DOI: https://doi.org/10.1038/nature13240

Kleffner, R., Flatten, J., Leaver-Fay, A., Baker, D., Siegel, JB., Khatib, F., & Cooper, S. 2017. Foldit Standalone: A video game-derived protein structure manipulation interface using Rosetta. *Bioinformatics*, 33(17): 2765–2767. DOI: https://doi.org/10.1093/bioinformatics/btx283

Klepac, P., Kissler, S., Gog, J. 2018. Contagion! The BBC Four Pandemic – The model behind the documentary. *Epidemics*, 24: 49–59. DOI: https://doi.org/10.1016/j.epidem.2018.03.003

Koepnick, B., Flatten, J., Husain, T., Ford, A., Silva, DA., Bick, MJ., Bauer, A., Liu, G., Ishida, Y., Boykov, A., Estep, R.D., Kleinfelter, S., Nørgård-Solano, T., Wei, L., Players, F., Monteline, G.T., DiMaio, F., Popović, Z., Khatib, F., Cooper, S., Baker, D. 2019. De novo protein design by citizen scientists. *Nature* 570: 390–394. DOI: https://doi.org/10.1038/s41586-019-1274-4

Krenn, M., Malik, M., Fickler, R., Lapkiewicz, R., Zeilinger, A. 2016. Automated Search for new Quantum Experiments. *Physical Review Letters*. 116(1): 090405. https://doi.org/10.1103/PhysRevLett.116.090405

Lake, BM., Salakhutdinov, R., Tenenbaum, J.B. 2015. Human-level concept learning through probabilistic program induction. *Science*, 350(6266): 1332–1338. DOI: https://doi.org/10.1126/science.aab3050

Lasecki, WS. 2019. On Facilitating Human-Computer Interaction via Hybrid Intelligence Systems. In *Proceedings of the 7th annual ACM Conference on Collective Intelligence.* ACM. Caraguatatuba, Brazil in October 2015, pp. 111-115. DOI: https://doi.org/10.1145/2857218.2857236

Lee, J., Kladwang, W., Lee, M., Cantu, D., Azizyan, M., Kim, H., … & Participants, E. 2014. RNA design rules from a massive open laboratory. *Proceedings of the National Academy of Sciences*, 111(6): 2122-2127. DOI: https://doi.org/10.1073/pnas.1313039111

Leoni, J., Tanelli, M., Strada, S.C., Berger-Wolf, T. 2020. Data-Driven Collaborative Intelligent System for Automatic Activities Monitoring of Wild Animals. In *2020 IEEE International Conference on Human-Machine Systems (ICHMS).* Rome, Italy on 7-9 September 2020, pp. 1-6. DOI: https://doi.org/10.1109/ICHMS49158.2020.9209350

Lévy, P., & Bononno, R. 1997. Collective intelligence: Mankind's emerging world in cyberspace. New York City, New York, USA: Perseus books.

Li, B., Hou, B., Yu, W., Lu, X., Yang, C. 2017. Applications of artificial intelligence in intelligent manufacturing: a review. Frontiers of Information Technology & Electronic Engineering, 18(1): 86-96. DOI: https://doi.org/10.1631/FITEE.1601885

Lim, CC., Kim, H., Vilcassim, M. Jr., Thurston, GD., Gordon, T., Chen, LC., Lee, K., Heimbinder, M., & Kim, SY. 2019. Mapping urban air quality using mobile sampling with low-cost sensors and machine learning in Seoul, South Korea. *Environment International*, 131:105022. DOI: https://doi.org/10.1016/j.envint.2019.105022

Lintott, C. 2019. The Crowd and the Cosmos: Adventures in the Zooniverse. Oxford. New York: Oxford University Press.

Lintott, CJ., Schawinski, K., Slosar, A., Land, K., Bamford, S., Thomas, D., Raddick, MJ., Nichol, RC., Szalay, A., Andreescu, D., Murray, P., Vandenberg, J. 2008. Galaxy Zoo: morphologies derived from visual inspection of galaxies from the Sloan Digital Sky Survey. *Monthly Notices of the Royal Astronomical Society*, 389(3): 1179–1189. DOI: https://doi.org/10.1111/j.1365-2966.2008.13689.x

Manson, E. (n.d.). Available at

https://edwardmanson.com/projects/bbc-pandemic-application-design/
https://edwardmanson.com/projects/bbc-pandemic-application-design/

Marcus, G. 2018. Deep Learning: A Critical Appraisal. *arXiv preprint*. arXiv:1801.00631.

Marcus, G. 2020. The Next Decade in AI: Four Steps Towards Robust Artificial Intelligence. *arXiv preprint.* URL: http://arxiv.org/abs/2002.06177

Masters, KL. and Galaxy Zoo Team. 2019. 'Twelve years of Galaxy Zoo', *Proceedings of the International Astronomical Union*, 14(S353): 205–212. DOI: 10.1017/S1743921319008615.

Michelucci, P. 2019. How Do We Create a Sustainable Thinking Economy?. *Medium*. November 4 2019. [online access at https://towardsdatascience.com/how-do-we-create-a-sustainable-thinking-economy-4d77839b031e last accessed 21 April 2021]

Michelucci, P., & Dickinson, J. L. 2016. The power of crowds. *Science*, 351(6268): 32-33.

Mitchell, M. 2011. Ubiquity symposium: Biological Computation. In *Ubiquity*. February 2011. Article 3. DOI: https://doi.org/10.1145/1940721.1944826

Noisetube. 2021. Available at https://play.google.com/store/apps/details?id=net.noisetube&hl=en_US&gl=US

Nugent, J., 2019. Game for good with Stall Catchers citizen science. *Science Scope*, 42: 10–11.

Nushi, B., Kamar, E., Horvitz, E. 2018. Towards Accountable AI: Hybrid Human-Machine Analyses for Characterizing System Failure. arXiv preprint. arXiv:1809.07424

Ormiston, R., Nguyen, T., Coughlin, M., Adhikari, RX., Katsavounidis, E. 2020. Noise Reduction in Gravitational-wave Data via Deep Learning. *Physical Review Research*, 2(3): 033066. DOI: https://doi.org/10.1103/PhysRevResearch.2.033066

Paul, K., Quinn, M.S., Huijser, MP., Graham, J., Broberg, L. 2014. An evaluation of a citizen science data collection program for recording wildlife observations along a highway. *Journal of Environmental Management*, 139: 180–187. DOI: https://doi.org/10.1016/j.jenvman.2014.02.018

Phylo. 2021. Available at https://phylo.cs.mcgill.ca/play.php

Posey, B. (n.d.). Smart Sensor. *IoT Agenda*. Available at https://internetofthingsagenda.techtarget.com/definition/smart-sensor (Last accessed 21 August 2020)

Prakash, N., Mathewson, KW. 2020. Conceptualization and Framework of Hybrid Intelligence Systems. *arXiv preprint.* arXiv:2012.06161

Quake Catcher (n.d.). Available at https://quakecatcher.net/sensor/

Rallapalli, G., Fraxinus Players, Saunders, DG., Yoshida, K., Edwards, A., Lugo, CA., Collin, S., Clavijo, B., Corpas, M., Swarbreck, D., Clark, M., Downie, JA., Kamoun, S., Team Cooper, MacLean, D. 2015. Lessons from Fraxinus, a crowd-sourced citizen science game in genomics. *eLife,* 4: e07460. DOI: https://doi.org/10.7554/eLife.07460

Russell, P. 1995. *The Global Brain Awakens: Our next evolutionary leap California*. Global Brain Inc.

Schäfer, T., & Kieslinger, B. 2016. Supporting emerging forms of citizen science: A plea for diversity, creativity and social innovation. *Journal of Science Communication*, 15(2): Y02.

DOI: https://doi.org/10.22323/2.15020402

Scribes of the Cairo Geniza. (n.d.). Scribes of the Cairo Geniza. Available at https://www.scribesofthecairogeniza.org/ (Last accessed 21 April 2021)

Scribes of the Cairo Geniza. 2021. Available at https://www.scribesofthecairogeniza.org/classify,

Segal, A., Gal, K., Kamar, E., Horvitz, E., & Miller, G. 2018. Optimizing Interventions via Offline Policy Evaluation: Studies in Citizen Science. In *Proceedings of the AAAI Conference on Artificial Intelligence*. New Orleans, Louisiana, USA on 2-7 February 2018.

Segal, A., Gal, K., Kamar, E., Horvitz, E., Miller, G. In press. Informed Intervention Design in Hybrid Intelligence Settings: Studies in Citizen Science.

Senior, A.W., Evans, R., Jumper, J., Kirkpatrick, J., Sifre, L., Green, T., Qin, C., Žídek, A., Nelson, A.W.R., Bridgland, A., Penedones, H., Petersen, S., Simonyan, K., Crossan, S., Kohli, P., Jones, D.T., Silver, D., Kavukcuoglu, K., Hassabis, D., 2020. Improved protein structure prediction using potentials from deep learning. *Nature* 577(7792): 706–710. DOI: https://doi.org/10.1038/s41586-019-1923-7

Silver, D., Huang, A., Maddison, C. J., Guez, A., Sifre, L., van den Driessche, G., Schrittwieser, J., Antonoglou, I., Panneershelvam, V., Lanctot, M., Dieleman, S., Grewe, D., Nham, J., Kalchbrenner, N., Sutskever, I., Lillicrap, T., Leach, M., Kavukcuoglu, K., Graepel, T., & Hassabis, D. 2016. Mastering the game of Go with deep neural networks and tree search. *Nature*, 529(7587): 484–489. DOI https://doi.org/10.1038/nature16961

Silver, D., Hubert, T., Schrittwieser, J., Antonoglou, I., Lai, M., Guez, A., Lanctot, M., Sifre, L., Kumaran, D., Graepel, T., Lillicrap, T., Simonyan, K., & Hassabis, D. 2018. A general reinforcement learning algorithm that masters chess, shogi, and Go through self-play. *Science,* 362(6419): 1140–1144. DOI: https://doi.org/10.1126/science.aar6404

Snik, F., Rietjens, JHH., Apituley, A., Volten, H., Mijling, B., Noia, AD., Heikamp, S., Heinsbroek, RC., Hasekamp, OP., Smit, JM., Vonk, J., Stam, DM., Harten, G. van, de Boer, J. & Keller, CU. 2014. Mapping atmospheric aerosols with a citizen science network of smartphone spectropolarimeters. *Geophysical Research Letters*, 41(20): 7351–7358. DOI: https://doi.org/10.1002/2014GL061462

Spatharioti, SE., Wylie, S., & Cooper, S. 2019. Using Q-Learning for Sequencing Level Difficulties in a Citizen Science Matching Game. *Extended Abstracts of the Annual Symposium on Computer-Human Interaction in Play Companion Extended Abstracts*. Barcelona Spain in October 2019, pp. 679–686. DOI: https://doi.org/10.1145/3341215.3356299

SpiderSpotter. 2021. Available at https://www.spinnenspotter.be/en/map

Stall Catchers. (n.d.) Join a global game that's trying to cure Alzheimer's. Available at https://stallcatchers.com (Last accessed 21 August 2020)

Stall Catchers. 2021. Available at https://stallcatchers.com/virtualMicroscope

Sterling, A. 2013. EyeWire Chat Commands. March 26 2013. Available at https://blog.eyewire.org/eyewire-chat-commands/ (Last accessed 21 April 2021)

Tinati, R., Luczak-Roesch, M., Simperl, E., Hall, W. 2017. An investigation of player motivations in Eyewire, a gamified citizen science project. *Computers in Human Behavior*, 73: 527–540. DOI: https://doi.org/10.1016/j.chb.2016.12.074

Tomašev, N., Paquet, U., Hassabis, D., & Kramnik, V. 2020. Assessing Game Balance with

AlphaZero: Exploring Alternative Rule Sets in Chess. *arXiv preprint*. URL: http://arxiv.org/abs/2009.04374

UKBMS (n.d.). Butterfly Transect Weekly Field Recording Form. Available at https://ukbms.org/sites/default/files/downloads/UKBMS%20F2%20Weekly%20recording%20form.pdf

van Horn, G., Mac Aodha, O., Song, Y., Cui, Y., Sun, C., Shepard, A., Adam, H., Perona, P., Belongie, S., 2018. The iNaturalist Species Classification and Detection Dataset. *arXiv preprint*. arXiv:1707.06642

Vepřek, LH., Seymour, P., & Michelucci, P. 2020. Human computation requires and enables a new approach to ethical review. *arXiv preprint*. arXiv:2011.10754.

Vinyals, O., Babuschkin, I., Czarnecki, WM., Mathieu, M., Dudzik, A., Chung, J., Choi, DH., Powell, R., Ewalds, T., Georgiev, P., Oh, J., Horgan, D., Kroiss, M., Danihelka, I., Huang, A., Sifre, L., Cai, T., Agapiou, JP., Jaderberg, M., … Silver, D. 2019. Grandmaster level in StarCraft II using multi-agent reinforcement learning. *Nature*, 575(7782): 350–354. DOI:https://doi.org/10.1038/s41586-019-1724-z

Walmsley, M., Smith, L., Lintott, C., Gal, Y., Bamford, S., Dickinson, H., Fortson, L., Kruk, S., Masters, K., Scarlata, C., Simmons, B., Smethurst, R., & Wright, D. 2020. Galaxy Zoo: Probabilistic Morphology through Bayesian CNNs and Active Learning. *Monthly Notices of the Royal Astronomical Society*, 491(2): 1554–1574. DOI: https://doi.org/10.1093/mnras/stz2816

Wang, J., Balaprakash, P., Kotamarthi, R. 2019. Fast domain-aware neural network emulation of a planetary boundary layer parameterization in a numerical weather forecast model. *Geoscientific Model Development*, 12: 4261–4274. DOI: https://doi.org/10.5194/gmd-12-4261-2019

Wiggins, A., Crowston, K. 2011. From Conservation to Crowdsourcing: A Typology of Citizen Science, In 2011 44th Hawaii International Conference on System Sciences. Kauai, Hawaii, USA on 4-7 January 2011,, pp. 1–10. DOI: https://doi.org/10.1109/HICSS.2011.207

Wilder, B., Horvitz, E., Kamar, E. 2020. Learning to Complement Humans. *arXiv preprint*. arXiv:2005.00582

Wintraecken, JJVR. 2012. The NIAM Information Analysis Method: Theory and Practice. Berlin, Germany: Springer Science & Business Media.

Woolley, AW., Chabris, C. F., Pentland, A., Hashmi, N., & Malone, TW. 2010. Evidence for a Collective Intelligence Factor in the Performance of Human Groups. *Science*, 330(6004): 686–688. DOI: https://doi.org/10.1126/science.1193147

Wootton, JR. 2017. Getting the public involved in Quantum Error Correction. *arXiv preprint*. arXiv:1712.09649

Wyler, D., Grey, F., Maes, K., & Fröhlich, J. 2016. Citizen science at universities: Trends, guidelines and recommendations. League of European Research Universities. http://www. leru. org/publications. League of European Research Universities. https://www.leru.org/files/Citizen-Science-at-Universities-Trends-Guidelines-and-Recommendations-Full-paper.pdf

Xue, Y., Davies, I., Fink, D., Wood, C., & Gomes, CP. 2016. Behavior Identification in Two-Stage Games for Incentivizing Citizen Science Exploration. In Rueher, M. (Ed.), *Principles and Practice of Constraint Programming*. Springer International Publishing. New York City, New York, USA. pp. 701–717. DOI: https://doi.org/10.1007/978-3-319-44953-1_44

Zanzotto, F.M. 2019. Viewpoint: Human-in-the-loop Artificial Intelligence. *Journal of Artificial Intelligence Research*, 64: 243–252. DOI: https://doi.org/10.1613/jair.1.11345